\documentclass[journal,twocolumn]{IEEEtran}

\usepackage{caption}
\usepackage{amssymb}
\usepackage{amsmath}
\usepackage{mathdots}
\usepackage{graphicx}
\usepackage{comment}
\usepackage{cite}
\usepackage{bbm}
\usepackage{float}
\usepackage{epstopdf}

\usepackage{color}

\usepackage{standalone}
\usepackage{tikz,pgfplots}
\usetikzlibrary{shapes}
\usetikzlibrary{intersections,backgrounds}
\usetikzlibrary{calc}

\renewenvironment{IEEEbiography}[1]
  {\IEEEbiographynophoto{#1}}
  {\endIEEEbiographynophoto}
\begin{document}

\title{Phase Noise Compensation for OFDM Systems}

\author{Amir Leshem and Michal Yemini
\thanks{
The phase noise compensation scheme presented in this paper is patented \cite{leshempatent}.

 The authors are with the Faculty of Engineering, Bar-Ilan University,
Ramat Gan 52900, Israel (email:
  leshem.amir2@gmail.com, michal.yemini.biu@gmail.com).

    This work was supported by the Israel
Science Foundation under Grant   903/2013.}

}

\date{}

\maketitle

\begin{abstract}
We describe a low complexity method for time domain compensation of phase noise in OFDM systems. We extend existing methods in several respects. First we suggest using the Karhunen-Lo\`{e}ve representation of the phase noise process to estimate the phase noise. We then derive an improved data-directed choice of basis elements for LS phase noise estimation and present its total least square counterpart problem. The proposed method helps overcome one of the major weaknesses  of OFDM systems.  We also generalize the time domain phase noise compensation to the multiuser MIMO context. Finally we present simulation results  using both simulated and measured phased noise. We  quantify the tracking performance in the presence of residual carrier offset.
\end{abstract}

\section{ Introduction}\label{sec:intro}

OFDM has become a prominent technology that is utilized in many modern communication systems including cellular systems such as 3GPP LTE \cite{cite3GPP}, wireless LAN (WLAN) \cite{citeIEEE802.11} and WiMax \cite{citeIEEE802.16}. In  future LTE systems (see 802.11ac) multi-carrier modulations such as   Generalized frequency division multiplex (GFDM),
Filter bank multi-carrier (FBMC),
Universal filtered multi-carrier (UFMC)
and Filtered OFDM (f-OFDM) \cite{6923528,6877912,7023145,7421450} may be implemented. In spite of its popularity and   robustness to multipath propagation, OFDM is known to be extremely affected by phase noise (PN) and frequency offset \cite{668721,658613,1044611,948268,4291833}.   MIMO OFDM receivers are very sensitive to the phase noise coming from the difference between the carrier frequency and the local oscillator (LO). In the case of very high data rates this is actually the limiting factor on performance.  High order modulations such as 256 QAM  are also severely affected by phase noise. Phase noise is typically modeled as a multiplicative noise. When the LO is locked to the carrier frequency, the phase noise is lower and is modeled as a finite power random process. When the LO is not locked the phase noise is modeled as a Wiener process with infinite power, the effect of such phase noise is analyzed from an information theoretic perspective in \cite{6151785}. In all typical 802.11 and LTE implementations the LO is locked; hence,
 we will concentrate on the first model. Phase noise can be considered to have two components: a common phase error (CPE) that is common to all carriers, and a time varying part that is frequency dependent. This part  is typically weaker than the CPE and generates the undesirable and harmful ICI.

A popular approach to ICI mitigation is the use of an MMSE equalizer in the frequency domain that balances the AWGN and the colored ICI. This approach can be found in many articles including \cite{1134136,1356210,4027596,4907430,4291833}. The alternative approach of PN cancellation/mitigation aims at jointly detecting the data and cancelling the PN as described in\cite{4533161,4156406}, or jointly estimating the channel and PN \cite{5262292,4568453,4567677,4810119,1677918} and even jointly estimating the channel, detecting the data and compensating for the PN as in \cite{7248499,4355336,6868950,7289484}.  These schemes suffer from high complexity that is impractical to implement at high rates and in high spectral efficiency communication systems. Yet a third  approach to reducing ICI is time domain processing. Casas et al. \cite{CasasBiracree2002} proposed a LS approach in the time domain where they represented the noise using one of two fixed bases: DFT or DCT. The dominant phase noise components are then estimated using LS fitting of a few basis vectors (typically the low frequency components due to phase noise properties). When a single basis vector is used, the method reduces to that of \cite{1134136,1356210,4027596}. The main drawback of the method in \cite{CasasBiracree2002} is that typically phase noise cannot be compactly represented in the fixed basis. Another LS PN compensation scheme estimates both the channel coefficients and the PN with assumed low complexity; however, the assumption that the number of pilot subcarriers  is larger than the number of transmitting antennas leads to high complexity in massive MIMO communication systems \cite{6375940,6415388,6736761}.
Another issue  we consider is  time variation of the statistical properties of the PN process.  Even when the LO is locked, the statistical properties of the PN are time varying due to external conditions such as temperature; e.g., heat from the mobile device,  thus making real time basis selection  very desirable for practical implementations. Therefore, statistical knowledge of the PN covariance matrix should be acquired.

Phase noise is  present in many communication systems such as WLAN, millimeter wave systems, full-duplex systems and massive MIMO systems. Full-duplex systems are especially affected by phase noise since  self-interference is performed in order to extract the received signals, see for example \cite{6488952,7637008,7815419}. The phase noise prevents transmitters from properly subtract their self-interference, this is especially harmful in high order modulations which demand high values of received SNR which are translated to low values of error vector magnitudes (EVM).
 OFDM millimeter wave systems are also affected by phase noise, see for example \cite{4457895,978061} ; here, as in other systems the effect of phase noise is most severe when using high order modulations, and is a key issue in implementing these systems.
Phase noise is also present in optical communication systems using coherent optical OFDM \cite{4488231}. In such networks the phase  is estimated digitally without using a
optical phase-locked loop, however, this estimation is not perfect and thus phase noise is cause this imperfect estimation.
The phase compensation scheme that we present is adaptable to the transmission of OFDM symbols in all of these systems.

In this paper we  replace the fixed basis proposed in previous works with
an adaptive basis which  is the best representation of the noise with respect to
the $L_2$ norm for a  random noise process  \cite{leshempatent}.
 Since in locked systems the phase noise behavior is quite stationary we can either pre-calibrate the phase noise PSD and then use an eigen-decomposition of the covariance matrix or estimate in real time the basis elements as well as the LS coefficients of the phase noise. The latter is more robust to environmental changes such as temperature which might affect the statistical properties of the phase noise.  We can also replace the LS estimation of the coefficient with a total least square estimation \cite{Markovsky20072283} that considers imperfection of the model.
 \\
The main contributions of this article are:
 1) Utilizing the Karhunen-Lo\`{e}ve  representation of the phase noise process covariance matrix as basis elements. This dramatically improves the results of \cite{CasasBiracree2002}.
 2) The introduction of the implementation of the total least square estimator  for phase noise mitigation schemes.
 3) We efficiently track of the subspace of the covariance matrix of the phase noise process for system with no information regarding the covariance matrix of the phase noise process.  This is performed utilizing the PAST algorithm \cite{Yang1995} and behaves well even in the presence of carrier frequency offset.
 4) We extended the above contributions for multiuser  uplink beamforming OFDM systems.

This paper is organized as follows: in section~\ref{sec:model} we present the phase noise model. Section~\ref{sec:comp_siso} presents our results  on phase noise compensation in SISO systems and   discusses the computational aspects of the compensation scheme. In Section~\ref{sec:enhancements} we discuss two enhancements to the compensation method. Section~\ref{sec:track} is dedicated to tracking the dominant subspace of the Karhunen-Lo\`{e}ve (KL) representation  presented in Section~\ref{sec:comp_siso}. In Section~\ref{sec:results_all} we present simulation and measured results of the proposed phase noise compensation method. Section~\ref{sec:num_res} covers simulated phase noise whereas Section~\ref{sec:mes_PN} discusses measured phase noise.  Section~\ref{sec:sim_track} analyzes simulations of the tracking algorithm
 proposed in Section~\ref{sec:track} for the measured phase noise.
Finally, Section~\ref{sec:conclisions}
concludes the paper.

\textit{Notations:} We denote  the convolution between two continuous time signals $x$ and $y$ at time $t$ by $x*y(t)$; for discrete time we  denote the convolution between the two discrete time signals $x$ and $y$ at time $k$ by $x*y(k)$.  Vectors and matrices appear in bold. Let $\boldsymbol a$ be a vector, we denote by $\boldsymbol a^T$ the transpose vector of $\boldsymbol a$ and by $\boldsymbol a^*$ the conjugate transpose of $\boldsymbol a$; note that if $\boldsymbol a$ is of dimension 1; i.e., scalar, then  $\boldsymbol a^*$ is equal to the conjugate of $\boldsymbol a$. Moreover, $\boldsymbol M^{\dag}$ is the Moore-Penrose pseudoinverse of a matrix $\boldsymbol M$. Finally, $\|\cdot\|$ denotes the $L_2$ norm and $\|\cdot\|_F$  denotes the Frobenius norm.

\section{ Phase noise model}\label{sec:model}

 In this section we  describe a mathematical model for the phase noise process and its effects on OFDM systems. Consider an OFDM system described by
\begin{flalign} \label{2.1)}
x(t)=\frac{1}{\sqrt{N}}\sum _{k=0}^{N-1}s(k)e^{j\omega _{k} t},  {\rm \; \; \; \; \; \; \; \; 0}\le {\rm t}\le {\rm T}_{{\rm s}}
\end{flalign}
 where $\omega _{k} =\omega _{0} +k\Delta \omega$ is the frequency of the $k$'th channel, $k=-\frac{N}{2} ,\ldots,\frac{N}{2} $, $\omega _{0}$ is the carrier frequency and
\begin{flalign} \label{2.2)}
\Delta \omega =\frac{2\pi }{T_{s} }
\end{flalign}
is the angular sampling frequency. Additionally, $s(k)$ is the symbol transmitted of the  $k$'th channel and is independent of symbols transmitted over other channels. The OFDM symbol passes through a time invariant channel (we  assume a quasi-stationary fading process) and the received signal $y(t)$ is given by
\begin{flalign} \label{2.3)}
y(t)=h*x(t)+n(t).
\end{flalign}

 Phase noise is multiplicative noise resulting from the jitter of the LO of the OFDM system. We can model the received signal with the effect of the phase noise as
\begin{flalign} \label{2.4)}
z(t)=y(t)e^{j\phi (t)},
\end{flalign}
where $\phi (t)$ is a random process that can be considered to be a filtered Gaussian process with PSD $P_{\phi } \left(f\right)$. The process $\psi \left(t\right)=e^{j\phi (t)}$ is the multiplicative noise process that can also include the residual frequency offset (a linear phase component) and the common phase error that is constant across frequencies. We want to estimate this and remove its effect, since it introduces inter-channel interference (ICI). We assume that $\psi \left(t\right)$ is a stationary process with a known  covariance $r_{\psi } \left(\tau \right)=E\left[\psi \left(t\right)\psi ^{*} \left(t-\tau \right)\right]$. This assumption is very reasonable when the LO is locked to a stable frequency source through a Phase-Locked Loop (PLL).

Let $\boldsymbol \psi =\left(\psi _{1} ,\ldots,\psi _{{\rm N} } \right)^{T}$ be a vector of N consecutive samples of the phase noise process $\psi _{m} =\psi \left(mT_{s} \right)$. We  define the covariance matrix of the phase noise process by
\begin{flalign} \label{2.5)}
\boldsymbol R_{\psi \psi } =\left[\begin{array}{ccc} {E\left(\psi _{1} \psi _{1}^{*} \right)} & \ldots & {E\left(\psi _{1} \psi _{N}^{*} \right)}\\
{\vdots} & \ddots\ & \vdots \\ {E\left(\psi _{N} \psi _{1}^{*} \right)} & \ldots & {E\left(\psi _{N} \psi _{N}^{*} \right)} \end{array}\right].
\end{flalign}

When trying to represent the phase noise along a single OFDM symbol it is natural to use the basis of the eigenvectors of $\boldsymbol R_{\psi \psi } $.
We decompose $\boldsymbol R_{\psi \psi }$ using an eigen-decomposition as
\begin{flalign} \label{2.6)}
\boldsymbol R_{\psi \psi } =\sum _{i=0}^{N-1}\mu _{i} \boldsymbol u_{i}\boldsymbol  u_{i}^{*},
\end{flalign}
where $\mu_0,\ldots,\mu_{N-1}$ denote the eigenvalues of $\boldsymbol R_{\psi \psi }$ and $\boldsymbol u_0,\ldots,\boldsymbol u_{N-1}$ denote their respective eigenvectors.

\noindent We now  describe the received signal and channel. We begin with a SISO model and then extend it to a SIMO model. Note that we are only interested in the SIMO case since the phase noise is identical on all spatial channels. We assume that the OFDM symbols are synchronized and that the cyclic prefix has been removed, so that the channel matrix can be  assumed to be circulant, and thus be  given by
\begin{flalign} \label{2.7)}
\boldsymbol H=\left[\begin{array}{ccccc} {h_{0} } & {h_{1} } & {\cdots } & {\cdots } & {h_{N-1} } \\ {h_{N-1} } & {h_{0} } & {\ddots } & {} & {h_{N-2} } \\ {\vdots } & {\ddots } & {\ddots } & {\ddots } & {\vdots } \\ {\vdots } & {} & {\ddots } & {\ddots } & {\vdots } \\ {h_{1} } & {\cdots } & {\cdots } & {h_{N-1} } & {h_{0} } \end{array}\right].
\end{flalign}
Furthermore, we  consider a single OFDM symbol. The time domain OFDM symbol is given by
\begin{flalign} \label{2.8)}
\boldsymbol x=\boldsymbol F_N^{*} \boldsymbol s
\end{flalign}
 where
\begin{flalign} \label{2.9)}
\boldsymbol s=\left[s_{0} ,\ldots,s_{N-1} \right]^{T}
\end{flalign}
is the frequency domain OFDM symbol, and
\begin{flalign}
\boldsymbol F_N=\frac{1}{\sqrt{N}}\begin{bmatrix}
  1     & 1     & \ldots & 1     \\
 1     & e^{-2 \pi \cdot 1 \cdot 1 /N}     & \ldots & e^{-2 \pi \cdot 1 \cdot (N-1)/N}     \\
 \vdots                   & \vdots                   & \ddots & \vdots                       \\
 1 & e^{-2\pi(N-1) \cdot 1/N} & \ldots & e^{-2\pi\cdot(N-1) \cdot (N-1)/N}
\end{bmatrix}
\end{flalign}
is the DFT matrix. Let  $\boldsymbol n=\left[n_{0} ,\ldots,n_{N-1} \right]$ be the additive white Gaussian noise. It follows that\begin{flalign} \label{2.13)}
\boldsymbol y= \boldsymbol H\boldsymbol F_N^{*} \boldsymbol s +\boldsymbol n=\boldsymbol H\boldsymbol x+\boldsymbol n
\end{flalign}
is the received signal when no phase noise is present.

The received OFDM symbol $\boldsymbol z=\left[z_{0} ,\ldots,z_{N-1} \right]^{T}$ is given by
\begin{flalign} \label{2.10)}
\boldsymbol z=e^{j\boldsymbol\Phi }  (\boldsymbol H\boldsymbol x+\boldsymbol n)
\end{flalign}
where
\begin{flalign} \label{2.11)}
\boldsymbol\Phi ={\rm diag}\left(\phi _{0} ,\ldots,\phi _{{\rm N} -1} \right)
\end{flalign}
is the phase noise vector.

\noindent Define a received data matrix $\boldsymbol Z$; this matrix is given by
\begin{flalign} \label{2.12)}
\boldsymbol Z={\rm diag}(\boldsymbol z)=\left[\begin{array}{ccc} {z_{0} } & {} & {} \\ {} & {\ddots } & {} \\ {} & {} & {z_{N-1} } \end{array}\right]=\boldsymbol Ye^{j\boldsymbol\Phi }
\end{flalign}

\noindent where
\begin{flalign} \label{2.14)}
\boldsymbol Y={\rm diag}(\boldsymbol y)=\left[\begin{array}{ccc} {y_{0} } & {} & {} \\ {} & {\ddots } & {} \\ {} & {} & {y_{N-1} } \end{array}\right].
\end{flalign}

Our problem is to estimate the phase noise and construct a time domain vector that cancels out the harmful effect of the phase noise.

\section{ Time domain compensation in SISO systems}\label{sec:comp_siso}

 We now describe a time domain method for reducing the phase noise. The idea is to use available pilot data  to estimate the coefficients of a representation of the phase noise. We first present the phase compensation algorithm in \cite{CasasBiracree2002}. Then we  discuss the choice of basis for the phase noise compensation. We show that using a fixed basis such as Fourier vectors or discrete cosine transform vectors does not yield large gains in terms of ICI cancellation. We then propose the use of data directed basis selection and show the improvement  achieved through this approach. This is important for ICI cancellation since common phase noise removal involves
 choosing the first basis vector to be the $N$ dimensional all ones vectors $\mathbbm{1}_{N} =\left[1,\ldots,1\right]^{T}$.

\noindent Let $\boldsymbol v_{0} ,\ldots,\boldsymbol v_{N-1}$ be a basis for ${\mathbb{C}}^{N} $. Denote the phase noise realization by $e^{j\boldsymbol \varphi } =\left[e^{j\phi _{0} } ,\ldots,e^{j\phi _{{\rm N} -1} } \right]^{T} $ and let $\boldsymbol \gamma =\left[\gamma _{0} ,\ldots,\gamma _{{\rm N} -1} \right]^{T} $ satisfy
\begin{flalign} \label{3.1)}
e^{-j\boldsymbol \varphi } =\sum _{k=0}^{N-1}\gamma _{k} \boldsymbol v_{k},
\end{flalign}
or equivalently
\begin{flalign} \label{3.2)}
e^{-j\boldsymbol \varphi } =\boldsymbol V\boldsymbol \gamma,
\end{flalign}
where $\boldsymbol V=\left[\boldsymbol v_{0} ,\ldots,\boldsymbol v_{N-1} \right]$. If we allow only $d$ basis vectors $\boldsymbol V^{(d)} =\left[\boldsymbol v_{0} ,\ldots,\boldsymbol v_{d-1} \right]$ we can pose the problem as a least squares problem. Had we known the vector $e^{-j\boldsymbol \varphi }$ our objective would have been finding $\hat{\boldsymbol \gamma }\in\mathbb{C}^N$ such that $\boldsymbol V^{(d)}\boldsymbol  \gamma $ cancels the phase noise optimally (in LS sense), i.e.
\begin{flalign} \label{ZEqnNum100752}
\hat{\boldsymbol \gamma }={\rm arg\; }\mathop{{\rm min}}\limits_{\boldsymbol \gamma } \left\| e^{-j\boldsymbol \varphi } -\boldsymbol V^{(d)} \boldsymbol \gamma \right\| ^{2}.
\end{flalign}
However, since $e^{-j\boldsymbol \varphi }$ is the vector which we want to estimate, we cannot use this naive approach. We discuss ways to overcome this issue below.
\subsection{LS compensation based on \cite{CasasBiracree2002}}\label{sec:LS_comp}
\begin{figure*}
\centering
\includegraphics[scale=0.5]{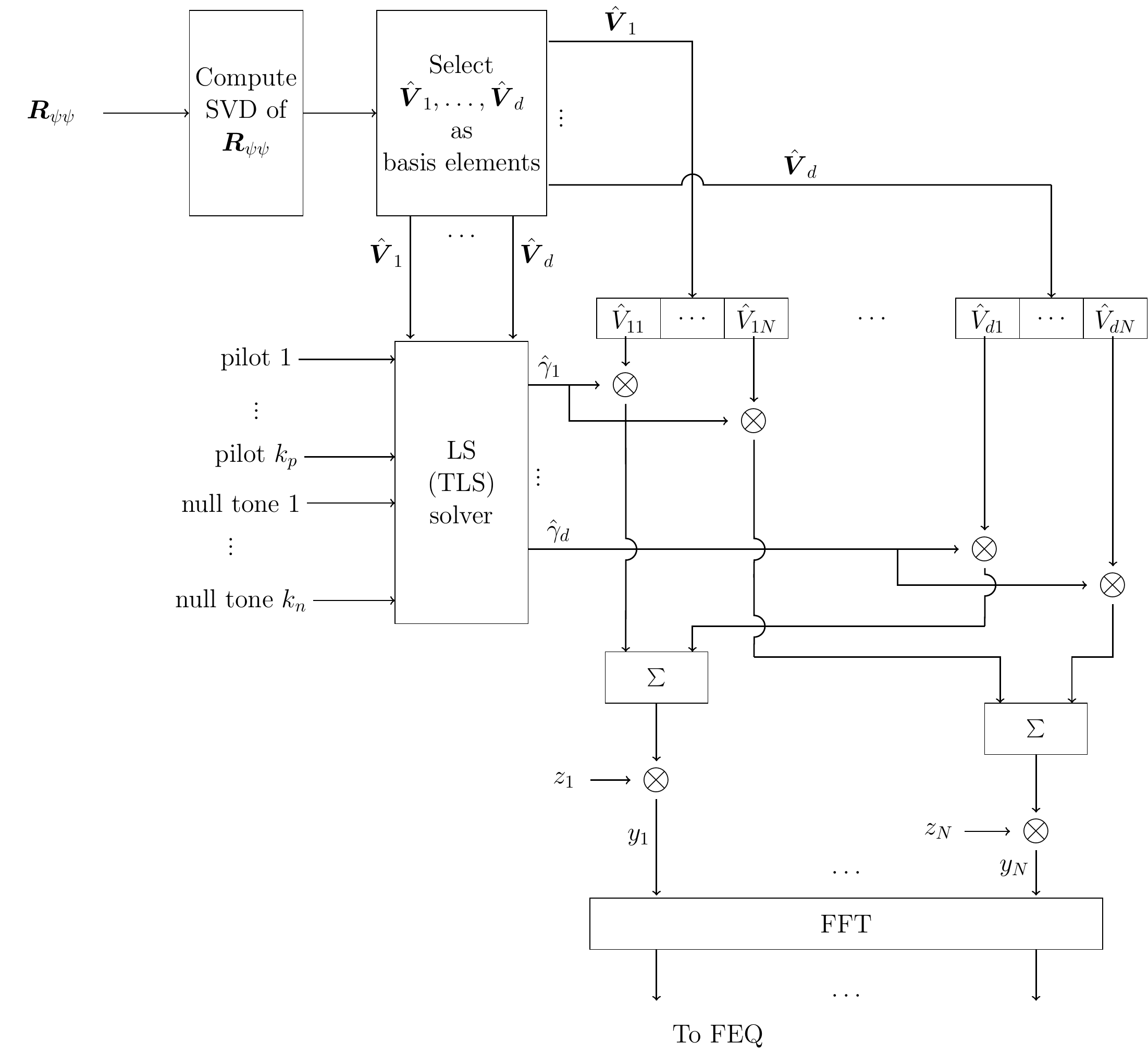}
\caption{Phase noise compensation process without subspace tracking.}
\label{Figure01}
\end{figure*}

 Since we do not know the phase noise $e^{-j\boldsymbol \varphi }$ we rely on known OFDM pilot tones. In this case we need to modify \eqref{ZEqnNum100752} assuming that we have known values $\boldsymbol s_{p} =\left[s_{i_{1} } ,\ldots,s_{i_{r} }\right]^{T}\in\mathbb{C}^{ n_{\text{pilot}}\times 1}$.  Let $\hat{\boldsymbol y}$ be an estimate of the time domain symbol with the phase noise removed:
\begin{flalign} \label{3.4)}
\hat{\boldsymbol y}=\boldsymbol Z\boldsymbol V^{(d)} \hat{\boldsymbol \gamma }\simeq \boldsymbol H\boldsymbol x+\boldsymbol n= \boldsymbol H\boldsymbol F_N^{*} \boldsymbol s +\boldsymbol n
\end{flalign}
Since $\boldsymbol H$ is diagonalized by the DFT matrix $\boldsymbol F_{N} $; i.e., $\boldsymbol H=\boldsymbol F_{N}^* \boldsymbol \Lambda \boldsymbol F_{N}$, we obtain that
\begin{flalign} \label{eq:3.5}
\hat{\boldsymbol s}=\boldsymbol \Lambda ^{-1} \boldsymbol F_{N} \boldsymbol Z\boldsymbol V^{(d)} \hat{\boldsymbol \gamma }.
\end{flalign}
is an estimate of the received OFDM frequency domain symbol. Defining
\begin{flalign} \label{3.6)}
\boldsymbol W=\boldsymbol \Lambda ^{-1} \boldsymbol F_{N} \boldsymbol Z\boldsymbol V^{(d)},
\end{flalign}
we obtain that our LS estimate of $\boldsymbol \gamma$ is given by
\begin{flalign}\label{ZEqnNum388909}
\hat{\boldsymbol \gamma }={\rm arg\; }\mathop{{\rm min}}\limits_{\boldsymbol \gamma } \left\| \boldsymbol s-\boldsymbol W\boldsymbol \gamma \right\| ^{2}.
\end{flalign}
Therefore, we obtain that
\begin{flalign} \label{3.8)}
\hat{\boldsymbol \gamma }=\boldsymbol W_{p} ^{\dag} \boldsymbol s_{p},
\end{flalign}
where $\boldsymbol W_{p} $ is obtained by choosing the rows that correspond to pilot tones alone.
The estimate of the phase noise cancellation vector is now given by
\begin{flalign} \label{3.9)}
e^{-j\boldsymbol \varphi } =\boldsymbol V\hat{\boldsymbol \gamma }.
\end{flalign}
Figure \ref{Figure01} depicts this phase noise compensation scheme.

Note that the components of $\boldsymbol Z$ are affected by noise and the noise is multiplied by $\boldsymbol \Lambda ^{-1} \boldsymbol F_{N}$. This suggests that the estimation of $\boldsymbol\gamma $ can be improved using  total least squares (TLS)  instead of the LS described above. We  discuss this in the next section.

\subsection{TLS compensation}\label{sec:TLS_comp}
As discussed above, in the training period, the training symbols $\boldsymbol s_{pilot} $ are sent over the channel. Since these training symbols are predefined and known by the receiver, our compensation problem can be represented  as a data least squares (DLS)  problem \cite{DeGroatDowling1993}. However, since we use the  basis $\boldsymbol V^{(d)}$ of size $d\leq N$ we  also consider discrepancies in $\boldsymbol s_p$.

 By Eq. (\ref{eq:3.5}) we have that
\begin{flalign}
\boldsymbol s\approx \boldsymbol \Lambda ^{-1} \boldsymbol F_{N} \boldsymbol Z\boldsymbol V^{(d)} \hat{\boldsymbol \gamma }.
\end{flalign}
The uncertainty in the model is caused  by the estimation of the channel matrix $\boldsymbol H,$  the additive noise of the channel and the reduced basis dimensions. It it therefore natural to consider the TLS estimation of $\boldsymbol \gamma$. Let $\Delta \boldsymbol W$ and $\Delta\boldsymbol s$ be such that
\begin{flalign}
(\boldsymbol W_{p}+\Delta \boldsymbol W)\boldsymbol\gamma=\boldsymbol s_{p}+\Delta\boldsymbol s
\end{flalign}
where as before, $\boldsymbol W_{p}$  is obtained by choosing the rows that correspond to pilot tones alone.

The  TLS problem is then
\begin{flalign}
&\arg\min_{\boldsymbol\gamma,\Delta \boldsymbol W,\Delta \boldsymbol s } \left\| [\Delta \boldsymbol W,\Delta \boldsymbol s] \right\|_F\nonumber\\
&\text{s.t.:} \hspace{1.5cm}(\boldsymbol W_{p}+\Delta \boldsymbol W)\boldsymbol\gamma=\boldsymbol s_{p}+\Delta\boldsymbol s
\end{flalign}
where as stated above (Section \ref{sec:intro}) the notation$\|\cdot\|_F$ denotes  the Frobenius norm.
The solution in terms of $\hat{\boldsymbol \gamma}$ is obtained by following Algorithm~1 in \cite{Markovsky20072283} which we  describe next.
Let $\boldsymbol W_p$ and $\boldsymbol s_p$ as stated above and let
\begin{flalign}\label{eq:SVD_prob}
[\boldsymbol W_p, \boldsymbol s_p]=\boldsymbol U\boldsymbol \Sigma \boldsymbol Q^T
\end{flalign}
be the singular value decomposition (SVD) of $[\boldsymbol W_p, \boldsymbol s_p]\in\mathbb{C}^{n_{\text{pilot}}\times (d+1)}$.
Denote
\begin{flalign}
\boldsymbol Q \triangleq\ \begin{bmatrix}
\boldsymbol q_{11} & \boldsymbol q_{12}\\
\boldsymbol q_{21} &  q_{22}
\end{bmatrix}
\end{flalign}
where $\boldsymbol q_{11}\in \mathbb{C}^{d\times d},\boldsymbol q_{12} \in \mathbb{C}^{d\times 1},\boldsymbol q_{21} \in \mathbb{C}^{1\times d}$ and $q_{22}\in \mathbb{C}$.
If $q_{22}\neq 0$ , note that $q_{22}$ is scalar, then
\begin{flalign}
\hat{\boldsymbol\gamma}_{\text{TLS}} = -\boldsymbol q_{12}/q_{22}.
\end{flalign}
When $q_{22}=0$, there is no solution and we set $\hat{\boldsymbol\gamma}_{\text{TLS}}=\hat{\boldsymbol\gamma}_{\text{LS}}$.
\newline
Note that in some practical implementations we trade off accuracy for lower complexity. However, if an SVD computation engine is available because of beamforming, for example, then this SVD engine can be used for solving the TLS problem in (\ref{eq:SVD_prob}).

\subsection{The basis vectors}

We now turn to the question of the choice of the basis vectors $\boldsymbol v_{0} ,\ldots,\boldsymbol v_{N-1}$. The authors of \cite{CasasBiracree2002} proposed using either the columns of the DFT matrix $\boldsymbol F_{N}$ or the columns of the DCT matrix. As will be seen in the simulations, this typically leads to minor improvement over simply cancelling the common phase but the ICI is still significant. We suggest a different approach and choose the basis elements using the properties of the phase noise process. We assume that the phase noise process has covariance
\begin{flalign} \label{3.10)}
\boldsymbol R_{\psi \psi } =\sum _{k=1}^{N}\mu _{k} \boldsymbol u_{k} \boldsymbol u_{k}^{*}
\end{flalign}
where $\boldsymbol u_{0} ,\ldots,\boldsymbol u_{N-1}$ are the eigenvectors corresponding to eigenvalues $\mu _{0} >\ldots>\mu _{{\rm N} -1}$, respectively.
 This basis is the best choice for representing random realizations of a random process with covariance $\boldsymbol R_{\psi \psi } $ (this is basically a KL representation of the process). Since the statistical properties of the phase noise process are stationary for quite long periods they can be calibrated in advance. ~

\subsection{Computational aspects}\label{comp_aspects}

\noindent The compensation of the phase noise involves $O((N\log N)\times d+ n_{\text{pilot}}\times d^{2})$  complex multiplications where $d$ is the number of the basis elements. This is due to the FFT complexity adding $d$ multiplications per symbol and the  additional complexity of the LS itself. We next explain these values in more detail.
Forming the matrix $\boldsymbol W$ involves computing
\begin{flalign} \label{5.1)}
\boldsymbol W=\boldsymbol \Lambda ^{-1} \boldsymbol F_{N} \boldsymbol Z \boldsymbol V^{(d)}.
\end{flalign}
Matrices $\boldsymbol \Lambda$ and $\boldsymbol Z$ are diagonal matrices and  matrix $\boldsymbol F_n$ is symmetric. Thus, calculating of the multiplication  $\boldsymbol Z \boldsymbol V^{(d)}$ requires at most $N\times d$ complex multiplication operations. Using the fast Fourier transform (FFT), we have that the complexity of  calculating $\boldsymbol F_{N} \boldsymbol Z \boldsymbol V^{(d)}$ is  $O((N\log N)\times d)$. The, additional multiplication by  $\boldsymbol \Lambda ^{-1}$ does not increase the complexity.

Solving the LS problem \eqref{ZEqnNum388909} only involves  matrices of size $n_{\text{pilot}} \times d$ resulting in $O( n_{\text{pilot}}\times d^{2} )$ operations.

It follows that the complexity depends on $d$ and $n_{\text{pilot}}$ and is $O((N\log N)\times d+ n_{\text{pilot}}\times d^{2})$.

We note that the complexity of calculating the eigenvectors of the covariance matrix is $O(N^3)$. This is done once and thus the complexity over time
vanishes. However, since phase noise statistics can slowly vary due to frequency offset variations, we also provide a tracking solution. In Section \ref{sec:track} we demonstrate that we can track the $d$ leading eigenvectors with a complexity of $O(Nd)$ operations per update.
We can conclude that the complexity of our compensation scheme is  $O((N\log N)\times d+ n_{\text{pilot}}\times d^{2})$ even if tracking the $d$ leading eigenvectors is performed.
This is on par with the complexity of the compensation scheme presented in \cite{CasasBiracree2002}  while providing superior results due to the better and adaptive basis selection.

\section{Extensions}\label{sec:enhancements}
We now describe two extensions to the proposed method. These extensions can contribute substantially to the performance of the proposed scheme by increasing the number of available equations for PN mitigation.

\subsection{Using null tones}
While channel estimation can only use  tones in which energy has been transmitted, null tones can also provide  information on ISI. This depends, of course, on the amount of adjacent channel suppression; however, when the adjacent channel suppression is good, the ICI can be estimated  based on the null tones as well.

\subsection{Phase estimation in MIMO system}\label{sec:MIM_PN_comp}

When a MIMO system is used, typically all transceiver chains use the same LO. Hence the phase noise can be jointly estimated based on the pilot symbols from all the receive antennas. This extra information substantially enhances the applicability of the proposed method and improves the quality of the LS fitting of the coefficients, especially in modern 802.11ac and massive MIMO systems.

Next we provide a detailed example for adapting the algorithm presented in this paper to multiuser uplink beamforming OFDM. The example assumes transmitter phase compensation \cite{5073397,7932407},  for this reason it focuses on compensating for the receiver phase noise. Under the assumption of proper phase noise compensation at the transmitters, the phase noise at the receiver is the dominating noise compared with the compensated phase noise at the transmitters.  \\ Suppose that a receiver with $N_r$ antennas serves $N_u$ single antenna users.
 Each user $k_u\in\{1,\ldots,N_u\}$ sends the vector of symbols $\boldsymbol s_{k_u}=[s_{k_u,0},\ldots,s_{k_u,N-1}]^T$, by aiming to transmit the signal
\begin{flalign}
x_{k_u}(t)=\frac{1}{\sqrt{N}}\sum_{k=0}^{N-1}s_{k_u,k}e^{j\omega_{k} t},\quad 0\leq t\leq T_s.
\end{flalign}
Denote by $\phi_{k_{u}}(t)$  the residual   phase noise  at transmitter $k_u$ after performing phase noise compensation at the transmitter. The transmitted signal with the residual phase noise is given by\footnote{Note that (\ref{eq:trans_MAC}) displays our assumption that adequate  phase noise compensation was performed by the transmitters, thus the remaining phase noise at  the transmitters is negligible.}
\begin{flalign}\label{eq:trans_MAC}
\tilde{x}_{k_u}(t)=e^{j\phi_{k_{u}}(t)}x_{k_u}(t)\approx x_{k_u}(t)
\end{flalign}

 Denote by $h_{k_{u},k_r}(t)$  the channel between user $k_{u}$ and the receiver antenna $k_r$. The receiving signal at antenna $k_r$ is
\begin{flalign}
y_{k_r}(t)=\sum_{k_{u}=1}^{N_u}h_{k_{u},k_r}*\tilde{x}_{k_u}(t)+n_{k_{r}}(t).
\end{flalign}
Assuming a common LO for all the receiving antennas  (which is  the case in many communication systems), we can model
the received signal with the effect of the phase noise at the receiver as
\begin{flalign}
z_{k_r}(t)=e^{j\phi(t)}y_{k_r}(t);
\end{flalign}
an example of multiple LOs at the receiving end is discussed for example in \cite{7054485}.
Since we discuss in this example systems with transmitter phase noise compensation, we can reasonably assume that $e^{j\phi (t)}$ is considerably larger than $e^{j\phi_{k_{u}}(t)}$.
Also,
as before, we assume that the OFDM symbols are synchronized and that the cyclic prefix
has been removed, so that the channel matrix can be assumed
to be circulant, and thus for every antenna the channel is given by
\begin{flalign}
&\boldsymbol H_{k_{u},k_{r}} =
\nonumber\\
&\left[\begin{array}{ccccc} {h_{k_{u},k_{r},0} } & {h_{k_{u},k_{r},1} } & {\cdots } & {\cdots } & {h_{k_{u},k_{r},N-1} } \\ {h_{k_{u},k_{r},N-1} } & {h_{k_{u},k_{r},0} } & {\ddots } & {} & {h_{k_{u},k_{r},N-2} } \\ {\vdots } & {\ddots } & {\ddots } & {\ddots } & {\vdots } \\ {\vdots } & {} & {\ddots } & {\ddots } & {\vdots } \\ {h_{k_{u},k_{r},1} } & {\cdots } & {\cdots } & {h_{k_{u},k_{r},N-1} } & {h_{k_{u},k_{r},0} } \end{array}\right].
\end{flalign}
Let  $\boldsymbol H_{k_r} = \left[\boldsymbol H_{1,k_{r}},\ldots, \boldsymbol H_{N_u,k_{r}} \right]$ and  $\boldsymbol x=[\boldsymbol x_1^T,\ldots,\boldsymbol x_{N_u}^T]^T$  where
\begin{flalign}
\boldsymbol x_{k_u} = \boldsymbol F_N^*\boldsymbol s_{k_u}.
\end{flalign}
Further, let
\begin{flalign}
\tilde{\boldsymbol x}_{k_u} = e^{j\boldsymbol\Phi_{k_u}} \boldsymbol x_{k_u}
\end{flalign}
where $ \boldsymbol\Phi_{k_u}= {\rm diag}\left(\phi _{k_u,0} ,\ldots,\phi _{k_u,N -1} \right)$ and $\phi _{k_u,m} = \phi_{k_{u}}(m T_s)$.
Let $\tilde{\boldsymbol x}=[\tilde{\boldsymbol x}_1^T,\ldots,\tilde{\boldsymbol x}_{N_u}^T]^T$, we have that
\begin{flalign}
\boldsymbol y_{k_{r}} &=  \sum_{k_u=1}^{N_u}\boldsymbol H_{k_{u},k_{r}}\tilde{\boldsymbol x}_{k_u}+\boldsymbol n_{k_r}\nonumber\\
&=\boldsymbol H_{k_r}\tilde{\boldsymbol x}+\boldsymbol n_{k_r}.
\end{flalign}
Defining $\boldsymbol\Phi$ as in (\ref{2.11)}), we write
\begin{flalign}
\boldsymbol z_{k_{r}} =e^{j\boldsymbol\Phi} \boldsymbol y_{k_{r}} =e^{j\boldsymbol\Phi}(\boldsymbol H_{k_{r}}\tilde{\boldsymbol x} + \boldsymbol n_{k_{r}}).
\end{flalign}
Now, let $\boldsymbol H = [\boldsymbol H_1^T,\ldots,\boldsymbol H_{N_r}^T]^T$ and  $\boldsymbol n = [\boldsymbol n_1^T,\ldots,\boldsymbol n_{N_r}^T]^T$;  define $\boldsymbol y$ and $\boldsymbol z$ as follows
\begin{flalign}
\boldsymbol y = \begin{bmatrix}
\boldsymbol y_1\\
\vdots\\
\boldsymbol y_{N_r}
\end{bmatrix}=
\begin{bmatrix}
\boldsymbol H_{1}\tilde{\boldsymbol x}+\boldsymbol n_{1}\\
\vdots\\
\boldsymbol H_{N_r}\tilde{\boldsymbol x}+\boldsymbol n_{N_r}
\end{bmatrix}
=\boldsymbol H\tilde{\boldsymbol x}+\boldsymbol n
\end{flalign}
\begin{flalign}
\boldsymbol z = \begin{bmatrix}
\boldsymbol z_1\\
\vdots\\
\boldsymbol z_{N_r}
\end{bmatrix}=
\begin{bmatrix}
\boldsymbol e^{j\boldsymbol\Phi}\boldsymbol y_{1}\\
\vdots\\
\boldsymbol e^{j\boldsymbol\Phi}\boldsymbol y_{N_r}
\end{bmatrix}
=(\boldsymbol I_{N_{r}}\otimes e^{j\boldsymbol\Phi})\boldsymbol y.
\end{flalign}
Denote
\begin{flalign}
\boldsymbol Y = \text{diag}(\boldsymbol y)
\end{flalign}
it follows that
\begin{flalign}
\boldsymbol Z \triangleq \text{diag}(\boldsymbol z) = \boldsymbol Y(\boldsymbol I_{N_{r}}\otimes e^{j\boldsymbol\Phi}),
\end{flalign}
where $\otimes$ is the  Kronecker product\footnote{We note that in Figure \ref{Figure01} and Figure \ref{Figure02} the symbol $\otimes$ denotes the scalar multiplication. However, this is a special case of the Kronecker product.}.
Let $\boldsymbol V$ be defined to be a basis of $\boldsymbol R_{\psi\psi}$ (see (\ref{2.5)})), and define $\boldsymbol \gamma$ as in (\ref{3.1)}) and (\ref{3.2)}). We can conclude that
\begin{flalign}
(\boldsymbol I_{N_{r}}\otimes e^{j\boldsymbol\Phi})(\boldsymbol I_{N_{r}}\otimes (\boldsymbol V\boldsymbol\gamma)) = \boldsymbol I_{N\cdot N_r}
\end{flalign}
Choosing a ZF beamforming matrix $\boldsymbol B$ and assuming that $N_r>N_u$, we have that
\begin{flalign}\label{45}
\boldsymbol s\approx
 (\boldsymbol I_{N_{u}}\otimes \boldsymbol F_N)\boldsymbol B\boldsymbol Z (\boldsymbol 1_{N_{r}}\otimes (\boldsymbol V\boldsymbol\gamma))
\end{flalign}
where $\boldsymbol 1_{N_r}$ is a column vector of ones of size $N_r$.
 Define
\begin{flalign}
\boldsymbol W =(\boldsymbol I_{N_{u}}\otimes \boldsymbol F_N)\boldsymbol B\boldsymbol Z (\boldsymbol I_{N_{r}}\otimes \boldsymbol V)(\boldsymbol 1_{N_{r}}\otimes \boldsymbol I_{N}).
\end{flalign}
Since
\begin{flalign}
 (\boldsymbol 1_{N_{r}}\otimes (\boldsymbol V\boldsymbol\gamma))= (\boldsymbol I_{N_{r}}\otimes \boldsymbol V)(\boldsymbol 1_{N_{r}}\otimes \boldsymbol I_{N})\boldsymbol \gamma
\end{flalign}
it follows  from  (\ref{45}) and from our assumption that adequate  phase noise compensation was performed by the transmitters
\begin{flalign}\boldsymbol s\approx
\boldsymbol W\boldsymbol \gamma.
\end{flalign}
We can obtain the linear  LS estimation by solving
\begin{flalign}\label{ZEqnNum3889092}
&\hat{\boldsymbol \gamma }={\rm arg\; }\mathop{{\rm min}}\limits_{\boldsymbol \gamma } \left\| \boldsymbol s-\boldsymbol W\boldsymbol \gamma \right\| ^{2}.
\end{flalign}
Therefore, as before, we obtain that
\begin{flalign} \label{3.8)}
\hat{\boldsymbol \gamma }=\boldsymbol W_{p} ^{\dag} \boldsymbol s_{p},
\end{flalign}
where $\boldsymbol W_{p} $ is obtained by choosing the rows that correspond to pilot tones alone of the different users. Alternatively, we can solve the respective TLS problem.

\section{Tracking the dominant subspace of the Karhunen Lo\`{e}ve representation}\label{sec:track}

There are several methods to obtain the correlation matrix $\boldsymbol R_{\psi \psi } $. The first is to pre-calibrate it and generate fixed basis vectors that are either measured or computed from the LO design. Hence, one can estimate $\boldsymbol R_{\psi \psi}$ from the data and apply an eigen-decomposition to obtain the basis vectors.
These alternatives are hard to implement since they depend on component variability in manufacturing and therefore needs to be performed for each chip separately. This might also  lead to performance degradation due to environmental changes such as temperature or vendor dependent behavior. Furthermore, residual carrier offset also affects the optimal basis. A better choice that makes it possible to overcome the non-stationarity of the phase noise process is to track a basis for the phase noise subspace. To track the phase noise vectors we propose using the PAST algorithm \cite{Yang1995}.
\begin{comment}
The are several methods to obtain the correlation matrix $\boldsymbol R_{\psi \psi} $ and to track its dominant signal subspace. The first is to pre-calibrate it and generate fixed basis vectors that are either measured or computed from the local oscillator design. These options are simple to implement but might lead to performance degradation due to environmental changes such as temperature or vendor dependent behavior. Hence one can estimate $\boldsymbol R_{\psi \psi} $ from the data and apply an eigen-decomposition in order to obtain the basis vectors. A better choice that allows overcoming non-stationarity of the phase noise process is to track a basis for the phase noise subspace. We propose to use the PAST algorithm to track the phase noise eigenvectors. Since we do not require that our basis elements will be orthogonal we do not need the deflation approach of PAST-d.
\end{comment}
Since we do not require  our basis elements to be orthogonal we do not need the deflation approach of PAST-d (see  \cite{Yang1995}). Alternatively, the subspace can be tracked using traditional methods such that appear in \cite{58320}.
The PAST algorithm is implemented as follows.

\subsection{The PAST algorithm for subspace tracking}
Let $\boldsymbol V^{(d)}_0$ be a matrix composed of the $d$ low frequency vectors of the DFT matrix or any other a-priori estimate of the phase noise dominant eigenvectors (i.e., eigenvectors of $\boldsymbol R_{\psi \psi } $ corresponding to higher eigenvalues). Let $\boldsymbol P_{0} =\boldsymbol I_{d\times d}$. At OFDM symbol $m$ we use $\boldsymbol V^{(d)}_{m-1}$ to perform the phase noise removal in the time domain as described above. The estimated symbols in the frequency domain $\hat {s}^{(m)} (k)$ are used to remove the desired signal from the received time domain signal and estimate the phase noise process at each time by
\begin{flalign} \label{5.2)}
\hat{\boldsymbol\varphi }_{m} (t)=e^{j\measuredangle \left(z_{m} (t)\left(\sum _{k=0}^{N-1}\hat{s}^{(m)} (k)\hat{h}^{(m)} (k)e^{j\omega _{k} t}  \right)^{-1} \right)}, 0\leq t\leq T_s.
\end{flalign}
$\boldsymbol R_{\psi \psi } $ is updated by
\begin{flalign} \label{5.3)}
\boldsymbol R_{\psi \psi }^{m} =(1-\alpha )\boldsymbol R_{\psi \psi }^{m-1} +\alpha\boldsymbol  \varphi _{m} \boldsymbol \varphi _{m}^{*}.
\end{flalign}
$\boldsymbol V^{(d)}_{m}$ is updated using the PAST algorithm as described in Table~\ref{table_PAST} (see \cite{Yang1995}).

\begin{table}
\centering
\begin{tabular}{|p{2in}|} \hline\\
\hspace{0.05cm} \small{Initialize} $\beta$,\\ [0.25cm]
$\begin{array}{l} {\boldsymbol y_{m} =(\boldsymbol V^{(d)}_{m-1})^{*} \hat{\boldsymbol \varphi }_{m} }, \\ [0.25cm]
{\boldsymbol h_{m} =\boldsymbol P_{m-1} \boldsymbol y_{m}, } \\[0.25cm]
{\boldsymbol g_{m} =\frac{1}{\beta +\boldsymbol y_{m}^{*} \boldsymbol h_{m} } \boldsymbol h_{m}, } \\ [0.25cm]
{\boldsymbol P_{m} =\frac{1}{\beta } \left (\boldsymbol P_{m-1} -\boldsymbol g_{m} \boldsymbol h_{m}^{*} \right),} \\ [0.25cm]
{\boldsymbol e_{m} =\boldsymbol x_{m} -\boldsymbol V^{(d)}_{m-1} \boldsymbol y_{m}, } \\ [0.25cm]
{\boldsymbol V^{(d)}_{m} =\boldsymbol V^{(d)}_{m-1} +\boldsymbol e_{m} \boldsymbol g_{m}^{*}. } \end{array}$ \\ \\\hline
\end{tabular}
\caption{PAST algorithm for subspace tracking.}
\label{table_PAST}
\end{table}

\begin{figure*}
\centering
\includegraphics[scale=0.5]{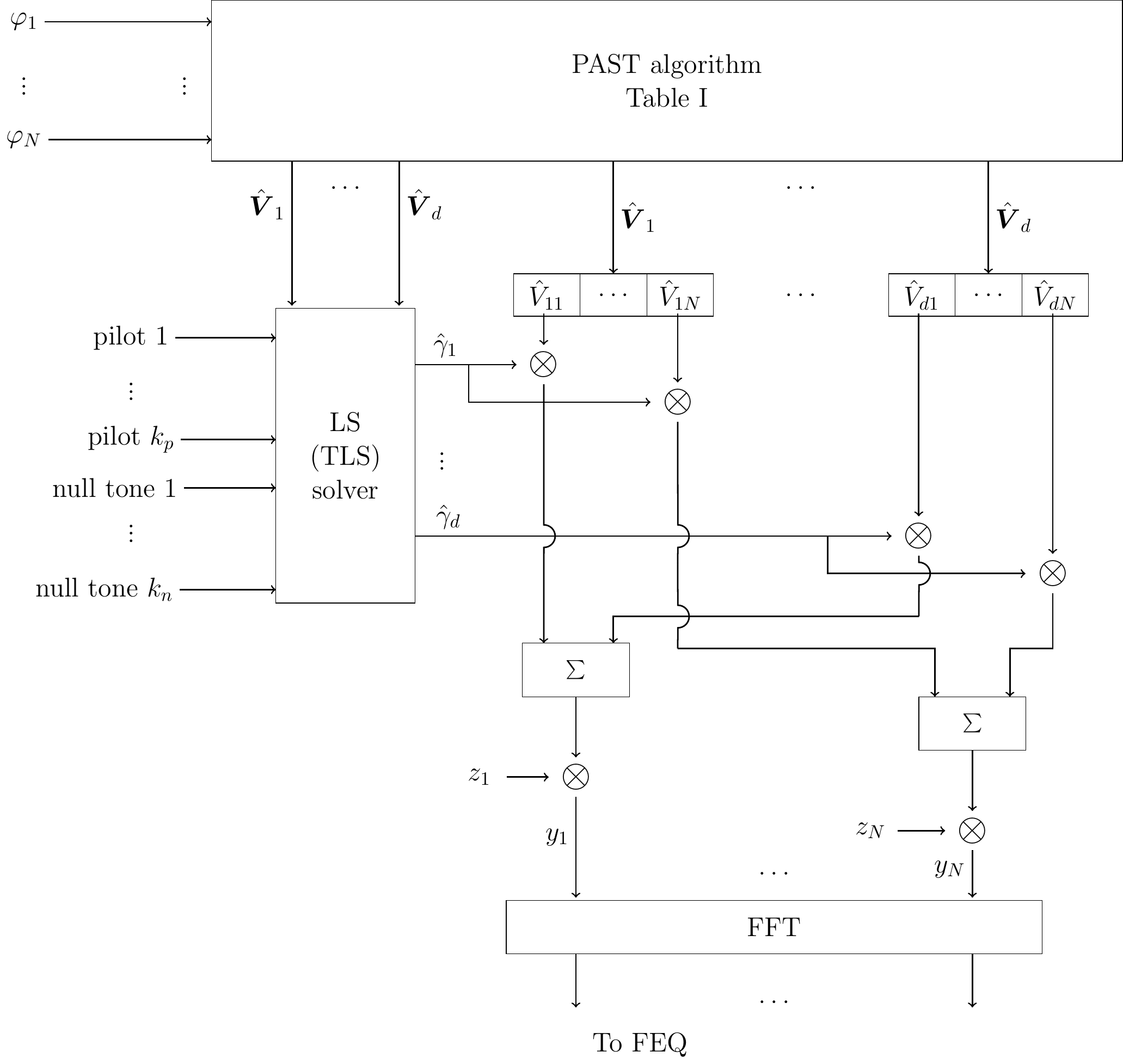}
\caption{Phase noise compensation process utilizing the PAST algorithm.}
\label{Figure02}
\end{figure*}

Note that  to track the subspace we do not need to compute $\boldsymbol R_{\psi \psi}$ and that is optional.  Figure \ref{Figure02} depicts this phase noise compensation scheme with subspace tracking.

\subsection{Residual carrier offset}
In the presence of residual carrier offset there is another time varying multiplicative noise; namely, the slowly varying residual carrier. While for small values (e.g., $1$ ppm) the residual carrier does not affect the decoding, it does have a detrimental effect on the common phase error estimation since the phase is no longer fixed at the pilots. Interestingly, this residual carrier can be incorporated into the KLT representation of the phase noise process by replacing $\psi(t)$ with the random process $\tilde{\psi}(t)$
\begin{flalign}
\tilde{\psi}(t)&=e^{j2\pi\Delta ft}\psi(t)\nonumber\\
&=e^{j(\phi(t)+2\pi\Delta ft)}
\end{flalign}
and then tracking the subspace of $\boldsymbol R_{\tilde{\psi}\tilde{\psi}}$ instead of $\boldsymbol R_{\psi \psi}$. We note that $ \tilde{\psi}(mT_s)$ and $ \tilde{\boldsymbol\psi}$ are defined similarly to the scenario in which there is no  residual carrier offset; that is, $\tilde{\psi}_m = \tilde{\psi}(mT_s)$ and $\tilde{\boldsymbol\psi} = [\tilde{\psi}_1,\ldots,\tilde{\psi}_N]^T$.
Let $c(t)= e^{j2\pi\Delta ft}$ and $c_m=c(mT_s)$. We can represent $\boldsymbol R_{\tilde{\psi}\tilde{\psi}}$ by the following identity
\begin{flalign}
\boldsymbol R_{\tilde{\psi}\tilde{\psi}} = \text{diag}\left(\boldsymbol c\right)\boldsymbol R_{\psi\psi}\text{diag}\left( \boldsymbol c\right)^*
\end{flalign}
where
\begin{flalign}
 \boldsymbol c= [c_1,\ldots,c_N]^T.
\end{flalign}
Let $\lambda$ be an eigenvalue of $\boldsymbol R_{\psi\psi}$ and $\boldsymbol v$ its respective eigenvector. By the definition of $\boldsymbol c$, $\text{diag}\left(\boldsymbol c\right)^*=\text{diag}\left(\boldsymbol c\right)^{-1}$,  thus
\begin{flalign}
\boldsymbol R_{\tilde{\psi}\tilde{\psi}}\text{diag}\left(\boldsymbol c\right)\boldsymbol v& = \text{diag}\left(\boldsymbol c\right)\boldsymbol R_{\psi\psi}\text{diag}\left( \boldsymbol c\right)^*\text{diag}\left( \boldsymbol c\right)\boldsymbol v \nonumber\\
&= \text{diag}\left( \boldsymbol c\right)\boldsymbol R_{\psi\psi}\text{diag}\left( \boldsymbol c\right) ^{-1}\text{diag}\left(\boldsymbol c\right)\boldsymbol v \nonumber\\
&= \text{diag}\left( \boldsymbol c\right)\boldsymbol R_{\psi\psi}\boldsymbol v \nonumber\\
&= \lambda\text{diag}\left( \boldsymbol c\right)\boldsymbol v
\end{flalign}
and we conclude that $ \text{diag}\left(\boldsymbol c\right)\boldsymbol v$ is an eigenvector of $\boldsymbol R_{\tilde{\psi}\tilde{\psi}}$
for the eigenvalue $\lambda$.
It follows that the basis elements of  $\boldsymbol R_{ \psi \psi}$; i.e., the covariance matrix of the PN process without the residual frequency offset, are multiplied by exponentials of the form $c(t)=e^{j2\pi \Delta ft} $. This is especially appealing when we implement the adaptive tracking of the KLT basis elements, as will be demonstrated in the simulations.
Note that the subspace tracking algorithm, does not require $\Delta f$ to be known, but it is affected by it implicitly.

\subsection{The asymptotic behavior of the PAST algorithm}
The asymptotic behavior of the PAST and PASTd algorithms is discussed in \cite{550149,544206,Yang:1996:ACA} for real valued independent identically distributed (i.i.d.) Gaussian random vectors.
\\
Under the reasonable assumption that the random process  $\boldsymbol\psi$ is  $M$-dependent, we can sample   $\boldsymbol\psi$ every  $M+1$ samples. This will slow the convergence by a linear factor of $M+1$.  To use the results of\cite{550149,544206,Yang:1996:ACA} the problem can be represented as a real problem, using standard transformation. Let,
\begin{flalign}
\boldsymbol V_{\text{2rep}}=\begin{bmatrix}
\text{Re}(\boldsymbol V) & -\text{Im}(\boldsymbol V)\\ \text{Im}(\boldsymbol V) & \text{Re}(\boldsymbol V)
\end{bmatrix}
\end{flalign}
it follows that
\begin{flalign}
\begin{bmatrix}
\text{Re}(e^{-j\boldsymbol\varphi}) \\ \text{Im}(e^{-j\boldsymbol\varphi})
\end{bmatrix}=\boldsymbol V_{\text{2rep}}
\begin{bmatrix}
\text{Re}(\boldsymbol\gamma) \\ \text{Im}(\boldsymbol\gamma)
\end{bmatrix}
\end{flalign}
and we can track the subspace of the matrix $\boldsymbol V_{\text{2rep}}$.
Thus,
 assuming that the vectors $\boldsymbol\psi$ are circularly symmetric complex normal vector, we can utilize the results of  \cite{544206,Yang:1996:ACA} and have that under mild conditions, the subspace $\boldsymbol V^{(d)}_{m}$ converges to $\boldsymbol V^{(d)}$ with probability 1 as $m\rightarrow\infty$.

\section{Experimental Analysis}\label{sec:results_all}
In this section we present the simulated and experimental analysis of the proposed PN compensation scheme in sub-section~\ref{sec:LS_comp}. This section is divided into three parts: Section~\ref{sec:num_res} covers simulation tests of the PN compensation scheme with no tracking. Section~\ref{sec:mes_PN} is dedicated to the analysis of the compensation scheme  with no tracking  performed on the  measured PN. Last, Section~\ref{sec:sim_track} is dedicated to the analysis of the tracking algorithm that was proposed in Section~\ref{sec:track}. This analysis is carried out on the measured PN of Section~\ref{sec:mes_PN}.
A detailed description of the   measurements that were used to produce the figure  in this sections is included in   \cite{1326984}. The SNR was chosen such that the phase noise is the dominant noise, and limits the reception of 256 QAM. Still, a noise figure of 7 dB is quite high and even with this strong noise the phase noise is still dominant. Additionally,  PN compensation complicates the design of the receiver, therefore it is the most cost effective when the PN is the dominant noise  which  typically occurs when transmitting  high order constellations (i.e., high SNR) together with multi-antenna receiver.

\subsection{ Simulation tests}\label{sec:num_res}

\begin{figure}
\centering
\includegraphics[scale=0.49]{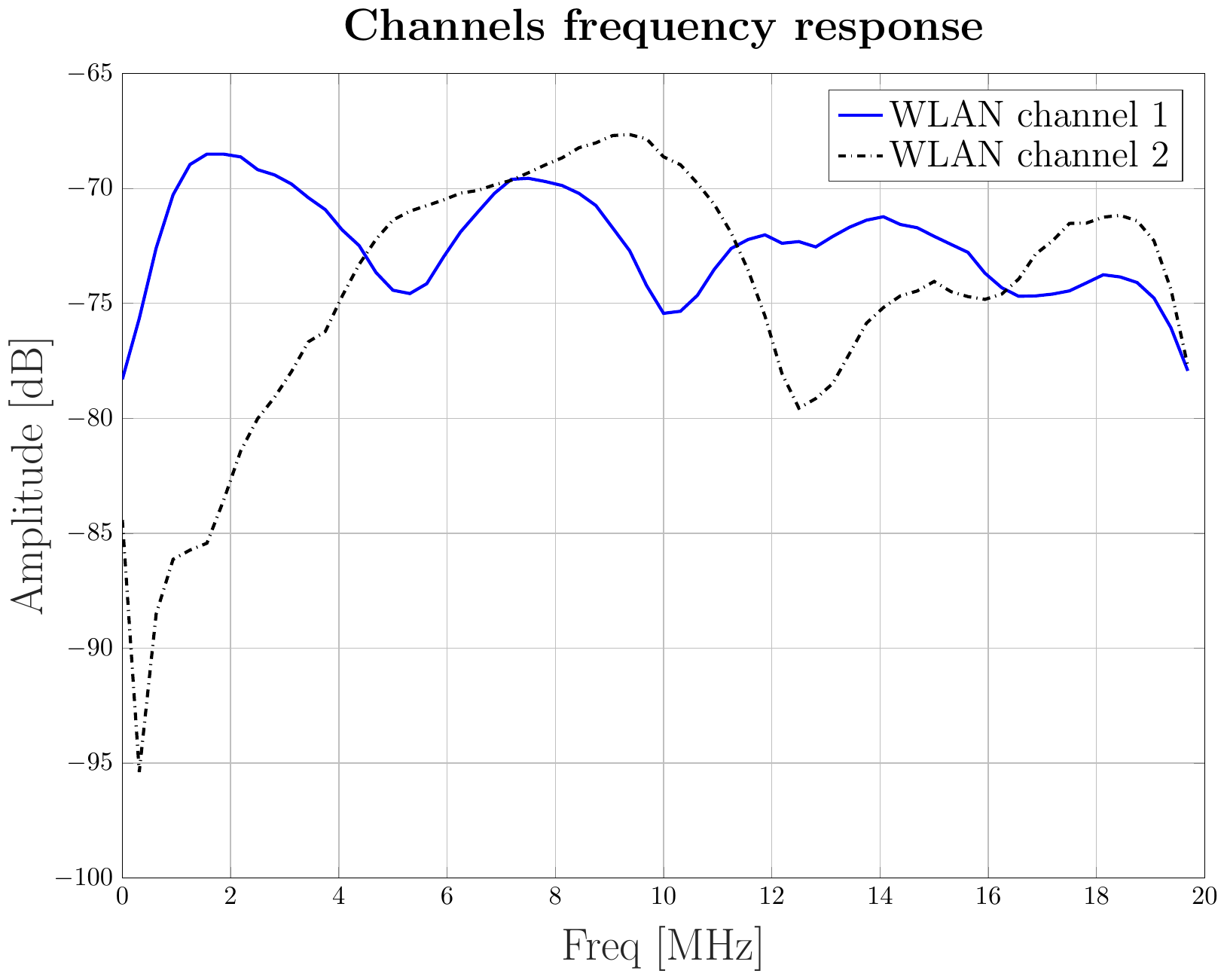}
\caption{Two $20$ MHz receive channels.}
\label{Figure03}
\end{figure}

\begin{figure}
  \centering
  % Requires \usepackage{graphicx}
 \includegraphics[scale=0.49]{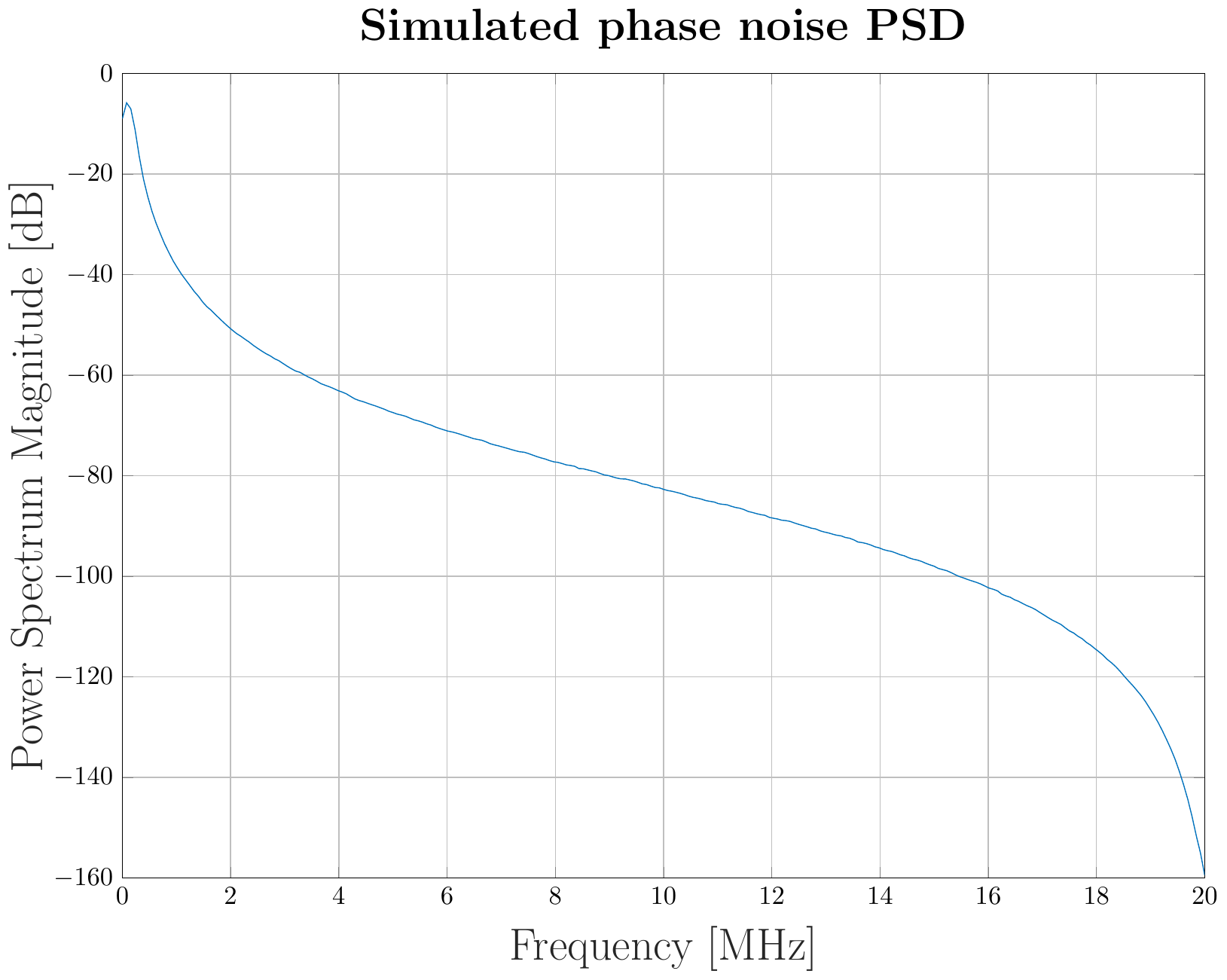}
  \caption{Simulated phase noise process PSD. $\sigma_{\phi}=3^{\circ}$.}
  \label{Figure04}
\end{figure}

We now present a simulated experiment testing the performance of the algorithm on real WLAN channels. This set of simulations assumes that the actual estimate of the eigenvectors of $\boldsymbol R_{\psi \psi}$ is given. An example of a measured channel is depicted in Figure~\ref{Figure03}. The transmitted power was 10 dBm, the assumed noise figure was 7 dB, the thermal noise was $-174\text{ dBm/Hz}$ and the bandwidth was $20\text{ MHz}$. The phase noise process  was generated using a second order Chebychev type I and a PSD of the phase noise process is depicted in Figure~\ref{Figure04}. The standard deviation of the phase noise was $\sigma_{\phi}=3^{\circ}$. At each time we  used  two receive channels and tone numbers $1-7, 21, 43, 58-64$ at each of the two receivers for pilot symbols $n_p=16$. The OFDM had $64$ tones; i.e., $N=64$ and the modulation was $256$ QAM.

\begin{figure}
  \centering
  % Requires \usepackage{graphicx}
 \includegraphics[scale=0.49]{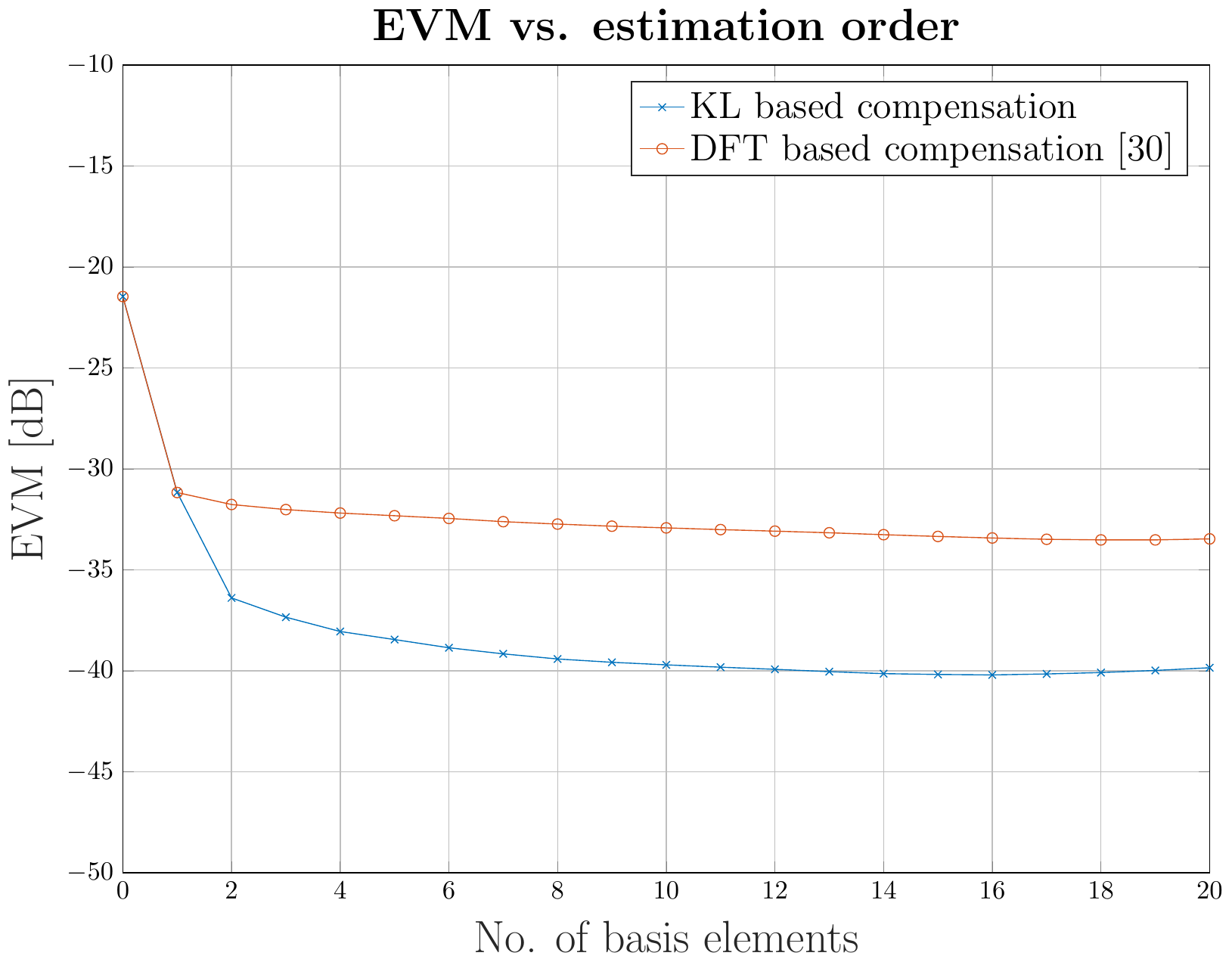}\\
  \caption{EVM vs. number of basis elements for DFT (which is the compensation scheme in  \cite{CasasBiracree2002}) and KL based methods. Phase noise $3^{\circ}$. $0$ - No compensation. $1$ - CPE compensation. $\sigma_{\phi} =3^{\circ}$}
  \label{Figure05}
\end{figure}

Figure~\ref{Figure05} depicts the dependence of the residual Error Vector Magnitude (EVM)  on the number of basis elements using KL basis vectors and  DFT basis vectors which the algorithm that was presented in \cite{CasasBiracree2002}.   The KL eigenvectors were computed based on 10 OFDM symbols. We averaged the EVM over 300 measured pairs of channels ($1\times 2$ systems, i.e., one transmitting antenna and two receiving antennas). For each channel we averaged the EVM over 100 OFDM symbols. Note that for a large number of basis elements, the EVM may be larger due to the insufficient number of equations when more basis elements are used than pilot symbols. The large gain of our compensation scheme compared  to the DFT basis presented in \cite{CasasBiracree2002} is clearly visible. Furthermore, the maximal tolerated EVM for 256 QAM for 5G is -32dB for a SISO transmission.  The  typical values for EVM are between -40dB and -35dB for MIMO channels. As can be seen, the method of  \cite{CasasBiracree2002}   narrowly meets the requirements  while our compensation scheme is well in the desired region.

\begin{figure}
  \centering
%  % Requires \usepackage{graphicx}
  \includegraphics[scale=0.49]{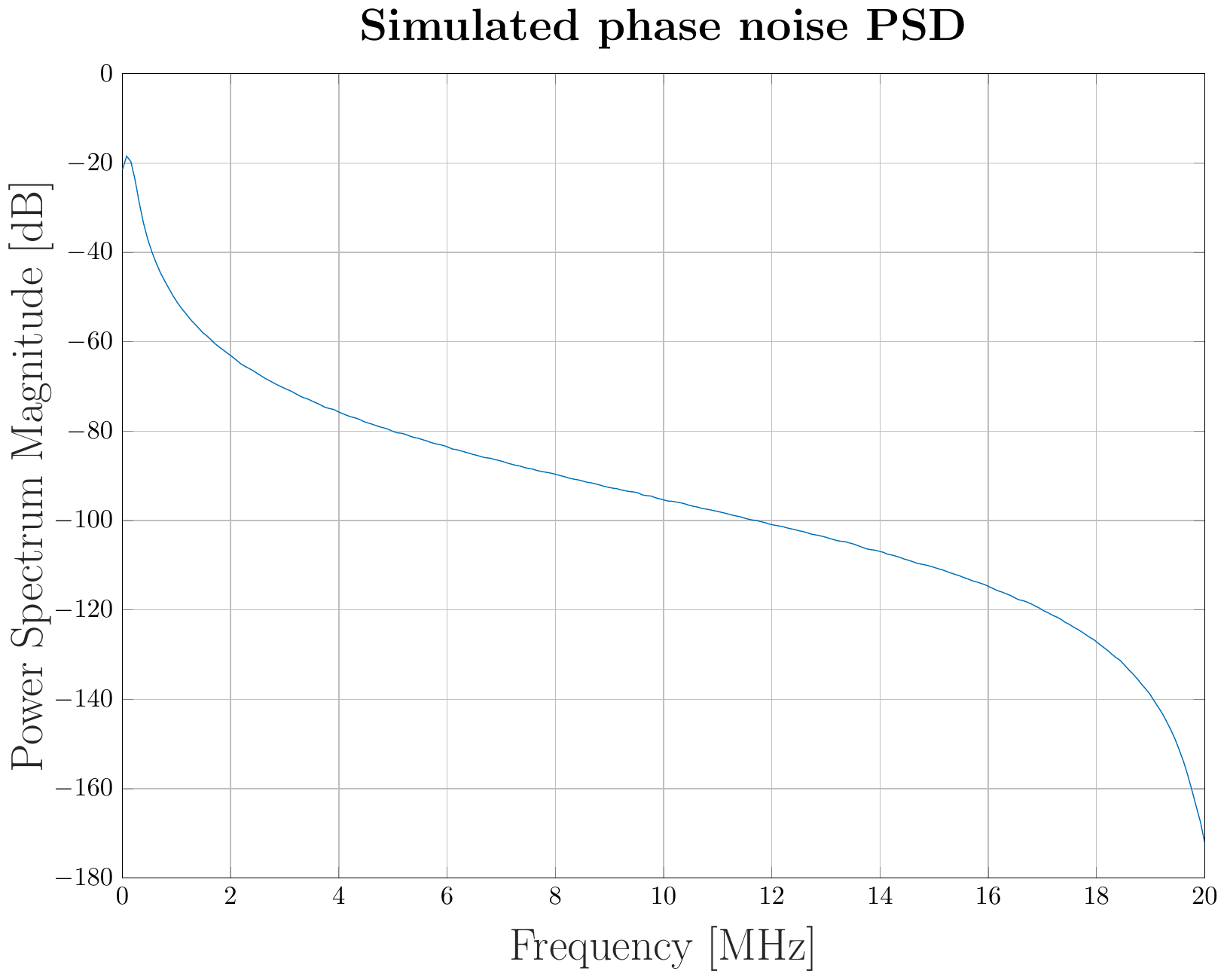}\\
  \caption{Simulated phase noise PSD. $\sigma_{\phi} =0.7^{\circ}$.}
  \label{Figure06}
  \end{figure}

\begin{figure}
  \centering
  % Requires \usepackage{graphicx}
 \includegraphics[scale=0.49]{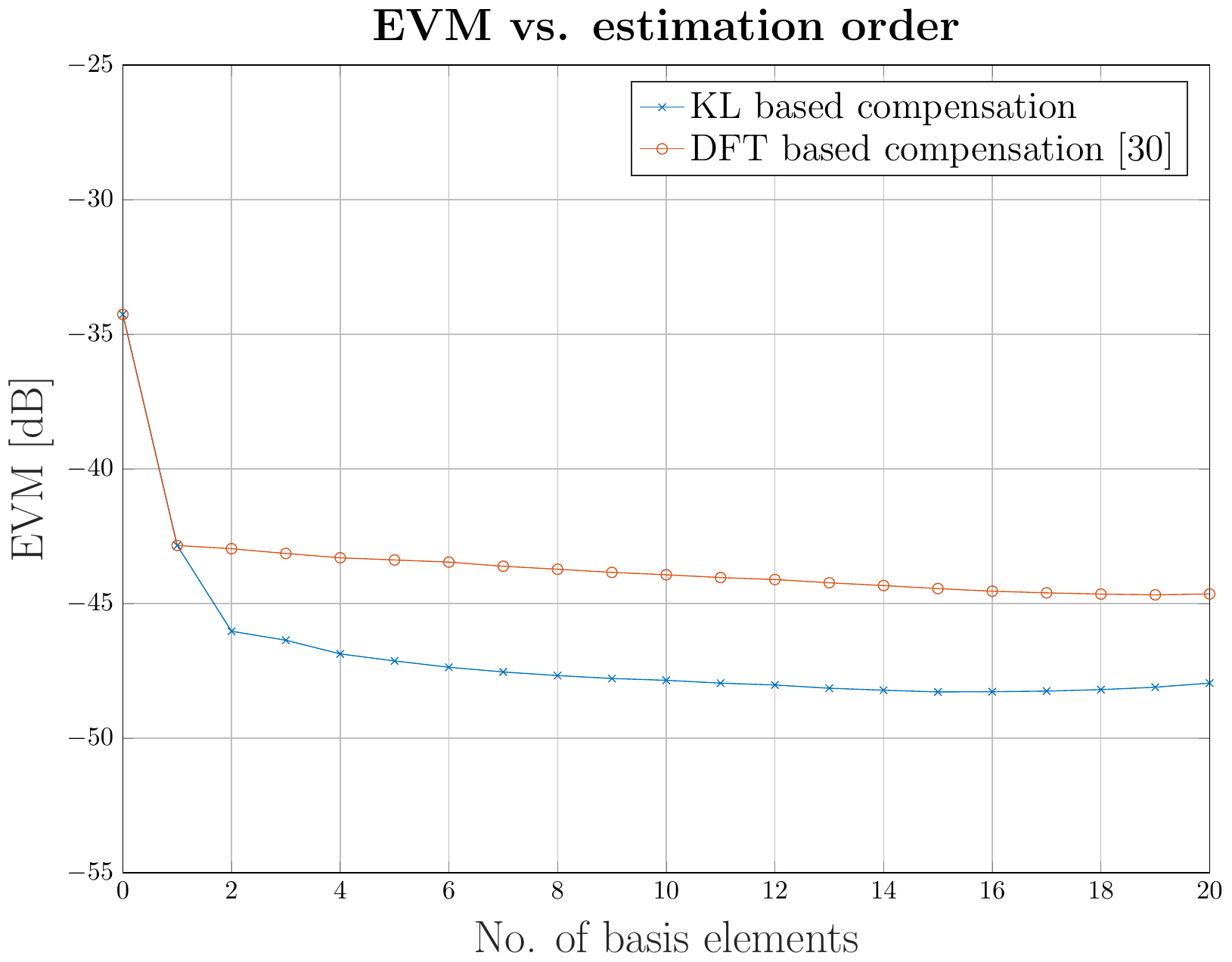}\\
  \caption{EVM vs. number of basis elements for DFT (see \cite{CasasBiracree2002}) and KL based methods. $0$ - No compensation. $1$ - CPE compensation. $\sigma_{\phi} =0.7^{\circ}$}
  \label{Figure07}
\end{figure}

To obtain  performance under good phase noise and channel conditions we  repeated the experiment with a simulated phase noise  with standard deviation of $\sigma_{\phi}=0.7^{\circ}$ and with the channel attenuation reduced by $5\text{dB}$ compared to Figure~\ref{Figure03}.
The results are presented in  Figures \ref{Figure06}-\ref{Figure07}. Even in this case there was a substantial gain achieved by cancelling the phase noise although the phase noise performance was reasonable even with CPE  compensation alone.

\begin{figure}
  \centering
  % Requires \usepackage{graphicx}
 \includegraphics[scale=0.49]{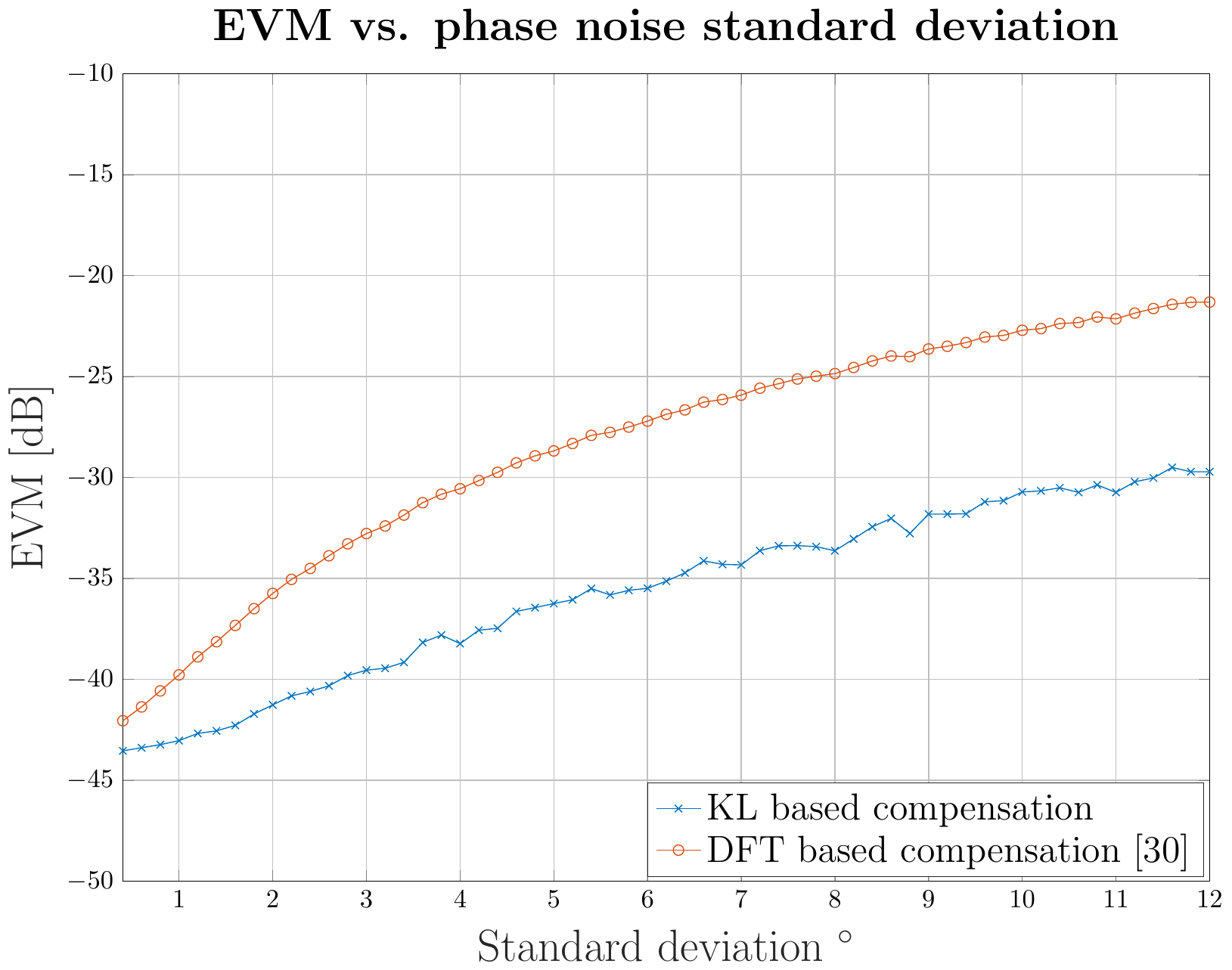}\\
  \caption{EVM vs. phase noise standard deviation. $8$ basis elements.}
  \label{Figure08}
\end{figure}

We also simulated the performance of our phase noise compensation scheme and that of Casas et al. \cite{CasasBiracree2002}, as a function of the of the standard deviation  of the phase noise.  That is, we chose a fixed number of 8 basis elements, and simulated the EVM as a function of the phase noise standard deviation. The results of this simulation are depicted in Figure \ref{Figure08}. It is clear that our scheme outperformed the scheme presented in \cite{CasasBiracree2002}. Specifically, using  DFT basis, a phase noise with standard deviation of $3^{\circ}$ is the maximal phase noise which can be tolerated for a constellation of 256QAM, whereas our scheme can tolerate up to $8^{\circ}$ standard deviation phase noise.
As before we used in this simulation 300 pairs of measured channels. For each pair channel we used 100 OFDM symbols; the phase noise process was produced for each of these channels as described at the beginning of Section \ref{sec:num_res}.

\begin{figure}
  \centering
  % Requires \usepackage{graphicx}
 \includegraphics[scale=0.49]{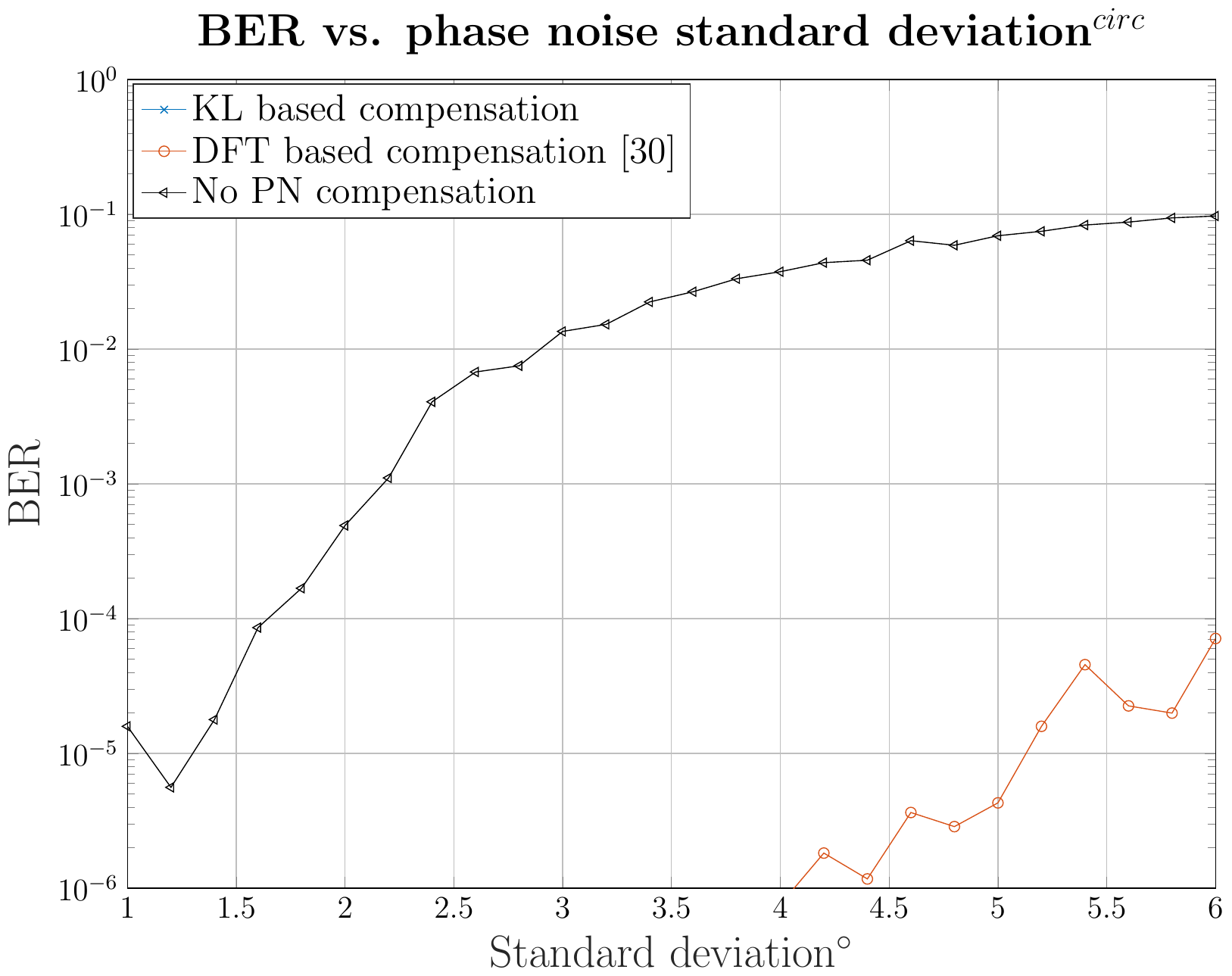}\\
  \caption{BER  vs. phase noise standard deviation. $8$ basis elements. }
  \label{Figure09}
\end{figure}

In addition, we performed  a coded simulation using a convolutional code with rate $\frac{1}{2}$, interleaver and soft-decision Viterbi decoding; its results are depicted in Figure~\ref{Figure09}. As before we used in this simulation 300 pairs of measured channels. For each pair channel we used 100 OFDM symbols; in total we used $7.68\times 10^6$ uncoded bits. The phase noise process was produced for each of these channels as described at the beginning of Section \ref{sec:num_res}.
As can be seen from Figure~\ref{Figure09} phase noise compensation is essential for correct decoding at the receiver. Further, the KL compensation scheme continued to outperform the DFT compensation scheme with bit error rate (BER) of less than $10^{-6}$.

\begin{figure}
\centering
\includegraphics[scale=0.49]{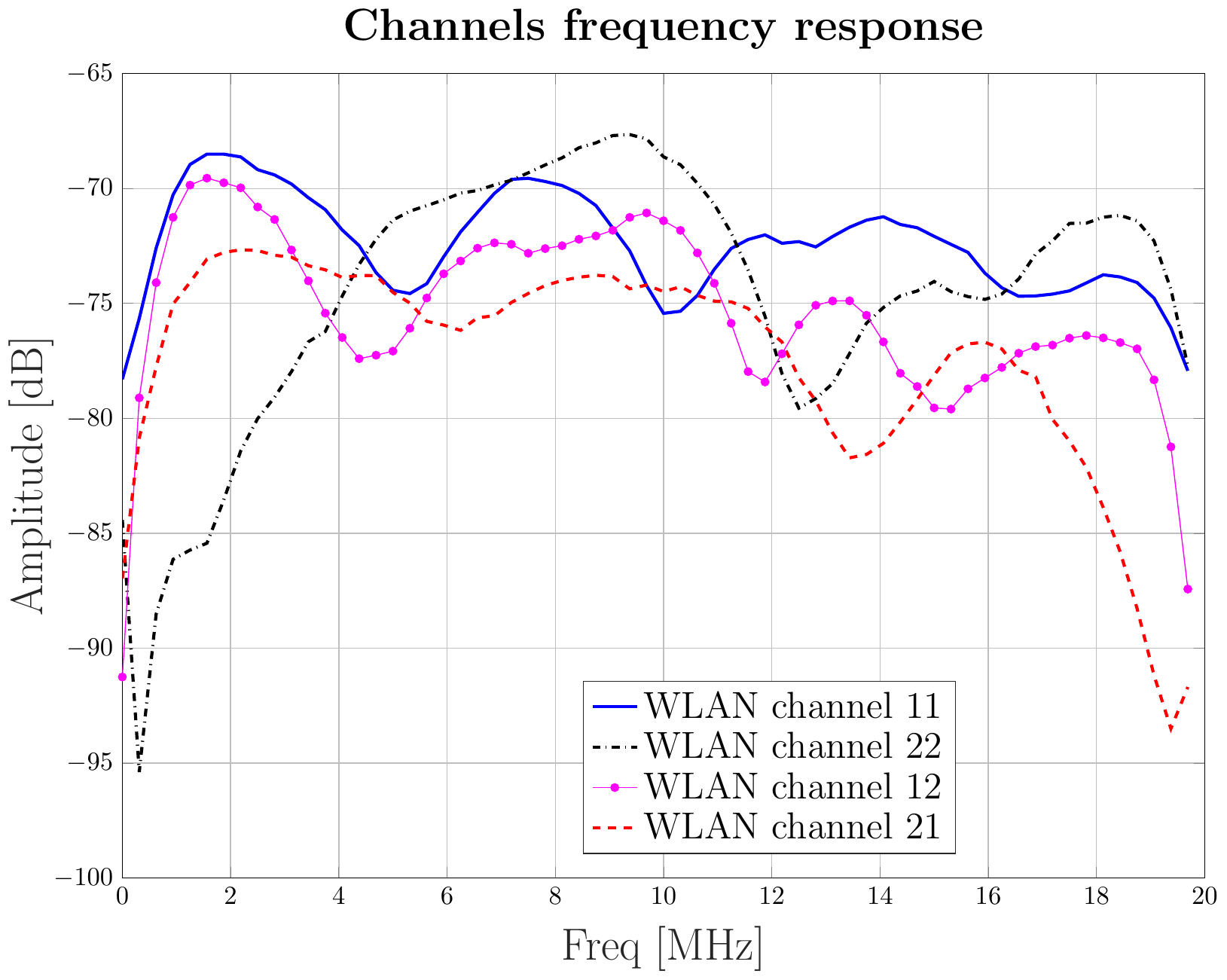}
\caption{Four $20$ MHz receive channels of $2\times 2$ communication system.}
\label{Figure10}
\end{figure}

\begin{figure}
  \centering
  % Requires \usepackage{graphicx}
 \includegraphics[scale=0.49]{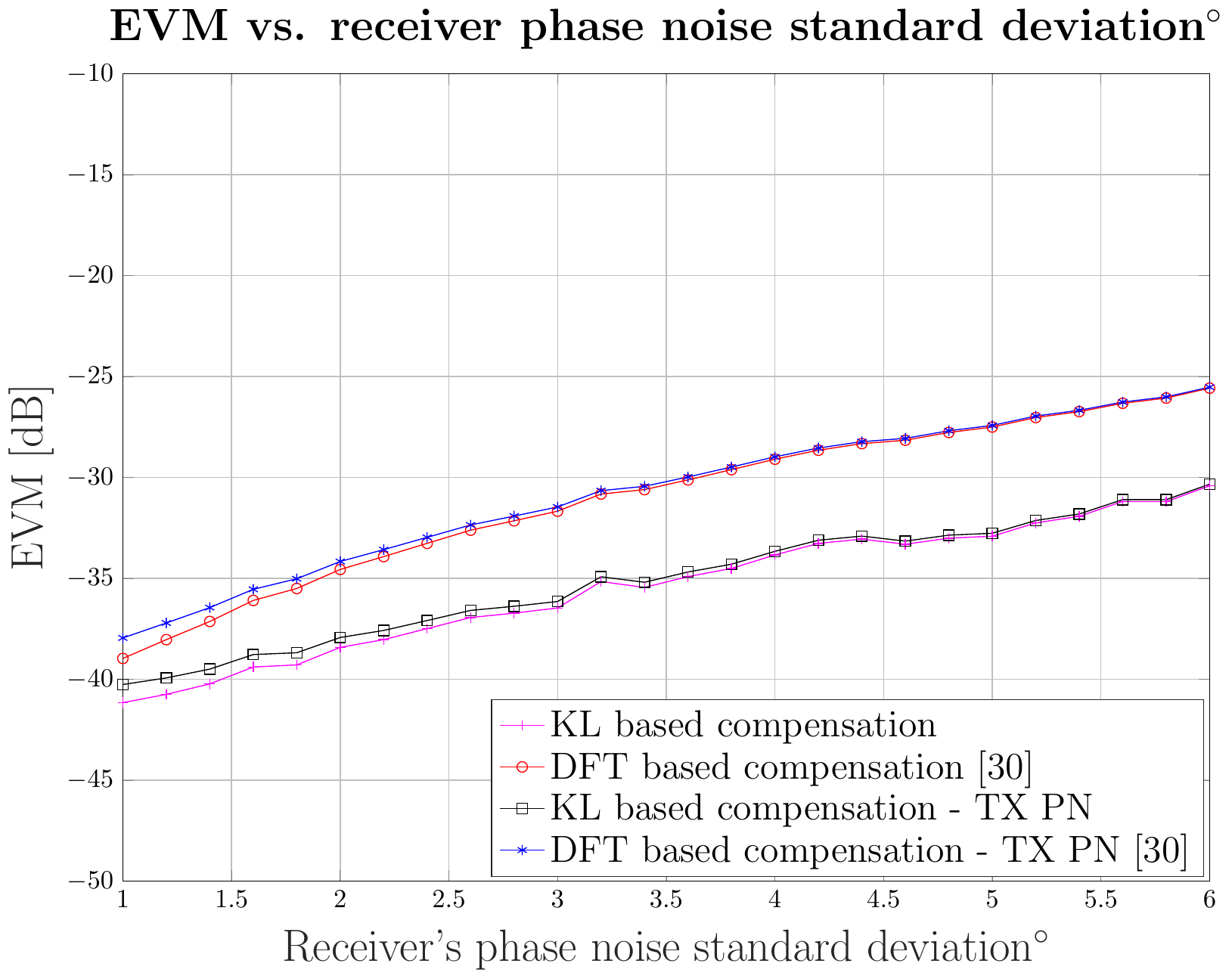}\\
  \caption{EVM  vs. phase noise standard deviation (at the receiver). $8$ basis elements. }
  \label{Figure11}
\end{figure}

The last simulation of this section depicts the performance of the phase noise compensation scheme for multi user MIMO system (see Section \ref{sec:MIM_PN_comp}) with two single antenna transmitters and two antenna receiver ($2\times 2$ system); Figure~\ref{Figure10} depicts an example of the channels of a $2\times 2$ communication system. The transmitted power of each transmitting antenna was 10 dBm. For each of the transmitting antennas we generated a phase noise process with standard deviation of $1^{\circ}$ which represents the residual phase noise  process at the transmitter.
The two phase noise processes at the transmitters were generated independently, and were also independent of the receiver phase noise process. We compared the performance of the system with and without transmitter phase noise. The results of the simulation are presented in Figure~\ref{Figure11}, the  receiver phase noise standard deviation was varied from $1^{\circ}$ to $6^{\circ}$. We can clearly see from Figure~\ref{Figure11} that when the  receiver phase noise standard deviation is greater than $1^{\circ}$, the performance of the phase noise compensation schemes with or without transmitter phase noise are very close to one another. Clearly, the method presented in this paper does not break down in the presence of independent residual transmitter phase noise.
Note that as before we used in this simulation 300 pairs of measured channels. For each pair channel we used 100 OFDM symbols; the receiver phase noise process was produced for each of these channels as described at the beginning of Section \ref{sec:num_res}.


\subsection{Measured phase noise analysis}\label{sec:mes_PN}

\begin{figure}
  \centering
  % Requires \usepackage{graphicx}
   \includegraphics[scale=0.49]{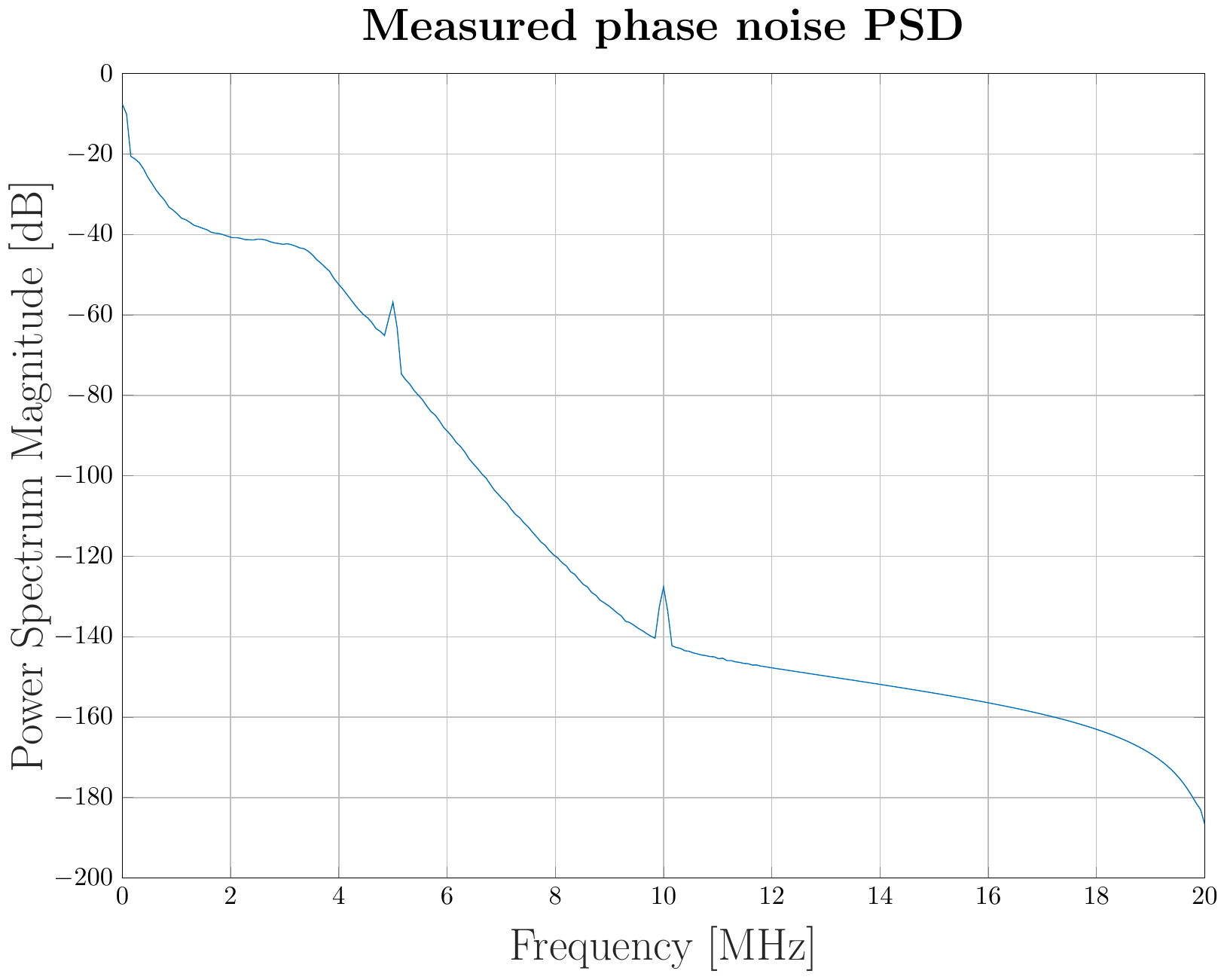}\\
  \caption[Caption for LOF]{PSD of measured phase noise.\footnotemark}
  \label{Figure12}
\end{figure}

\begin{figure}
  \centering
  % Requires \usepackage{graphicx}
 \includegraphics[scale=0.49]{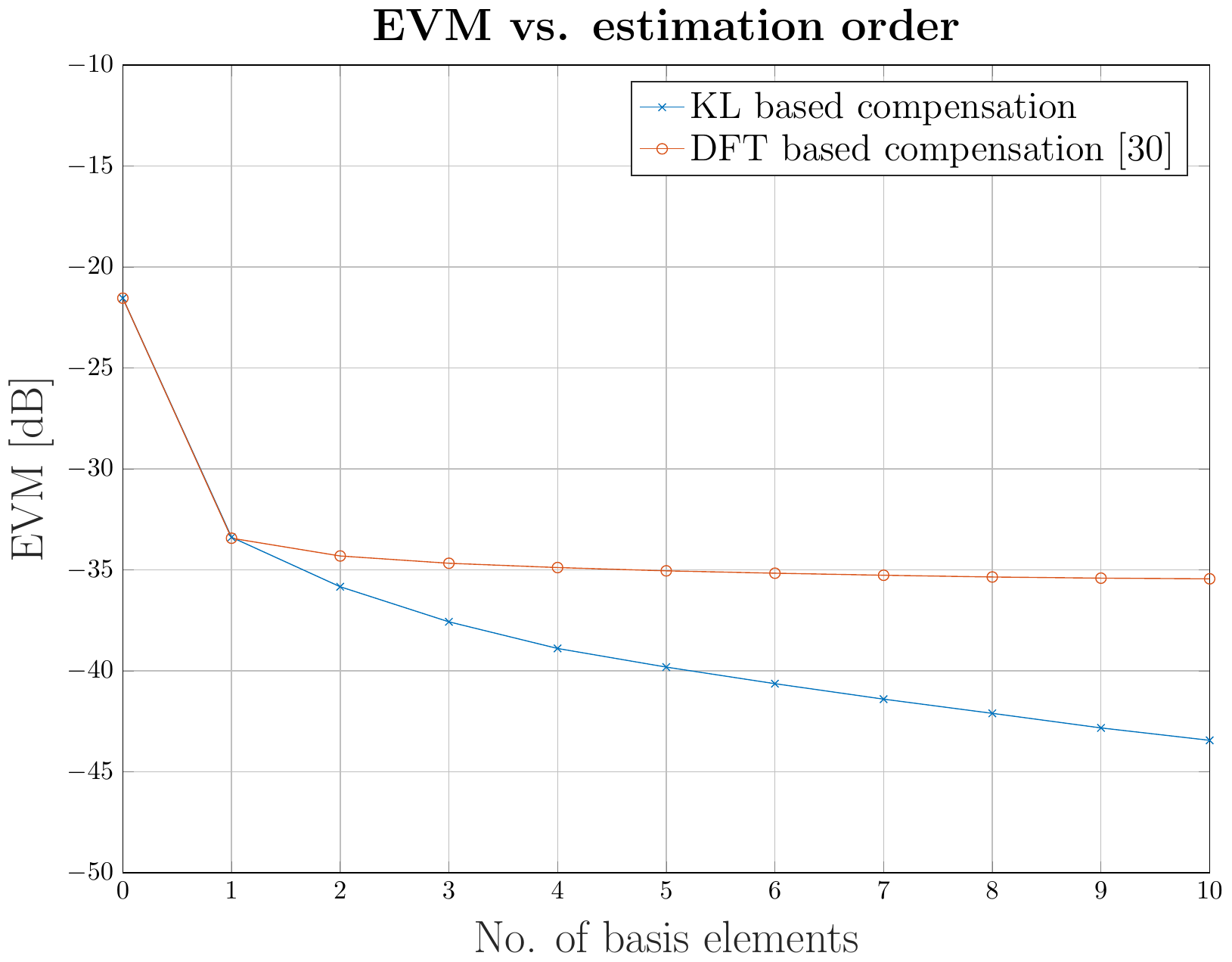}\\
  \caption{EVM vs. number of basis elements for DFT and KL based methods. No. of basis element: $0$ - No compensation. $1$ - CPE compensation. Measured phase noise.}
  \label{Figure13}
\end{figure}

In the second set of simulated experiments we  used measured phase noise samples. The measured signal comprised a sine wave at 5 MHz, sampled at 40 MHz and then filtered to remove the original sine wave of  5 MHz, thus  only the phase noise remained. The PSD of the phase noise is depicted in Figure~\ref{Figure12}. We  repeated the experiment above with samples of the measured phase noise. The results are depicted in Figure~\ref{Figure13}. There was a clear  gain of $5\text{dB}$ for $5$ basis elements and above.

\footnotetext{We note that the two spikes in the PSD of the measured phase noise process are due to residual of the sine
and its harmony of the sine wave at 5 MHz.}

\subsection{Simulations of the tracking algorithm for measured phase noise}\label{sec:sim_track}

\begin{figure}
  \centering
  % Requires \usepackage{graphicx}
  \includegraphics[scale=0.49]{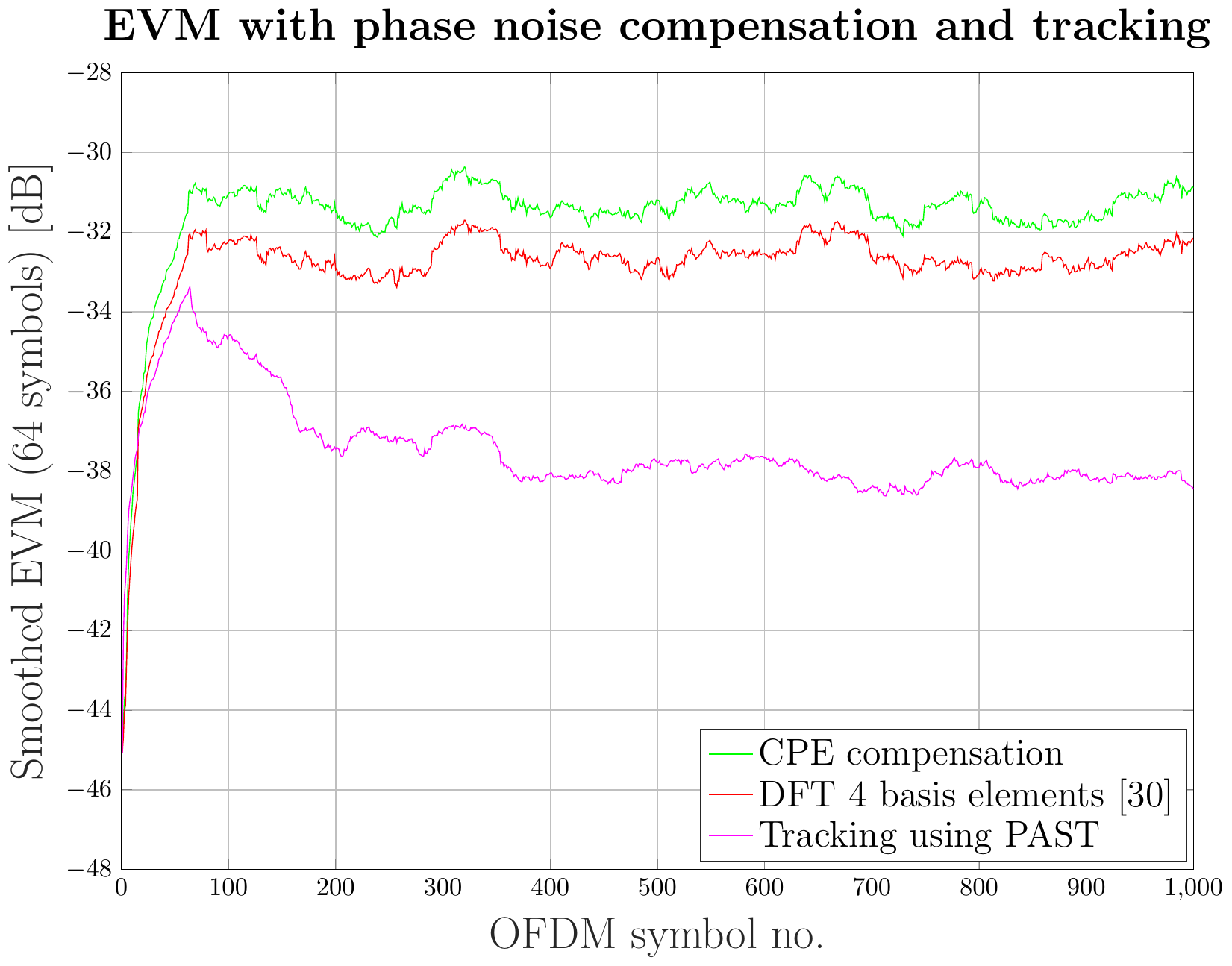}\\
  \caption{Tracking performance of PAST.  $\sigma_{\phi}=3.5^{\circ}$, residual carrier $1$ ppm, $4$  basis vectors.}
  \label{Figure14}.
\end{figure}

In this section we analyze the tracking capability of the PAST algorithm combined with decision direction. We used the same system as in previous sections with the measured phased noise (scaled to $3.5^{\circ}$ total phase noise) and 4 basis elements. However, we  also added a $1\text{ppm}$ residual carrier to model carrier offset for a carrier frequency of 5 GHz. It is clear (see Figure \ref{Figure14}) that our tracking scheme yields better phase and residual carrier compensation of 5dB, compared to the other  methods, i.e., the DFT of   \cite{CasasBiracree2002} and a simple CPE compensation.

\begin{figure}
  \centering
  % Requires \usepackage{graphicx}
  \includegraphics[scale=0.49]{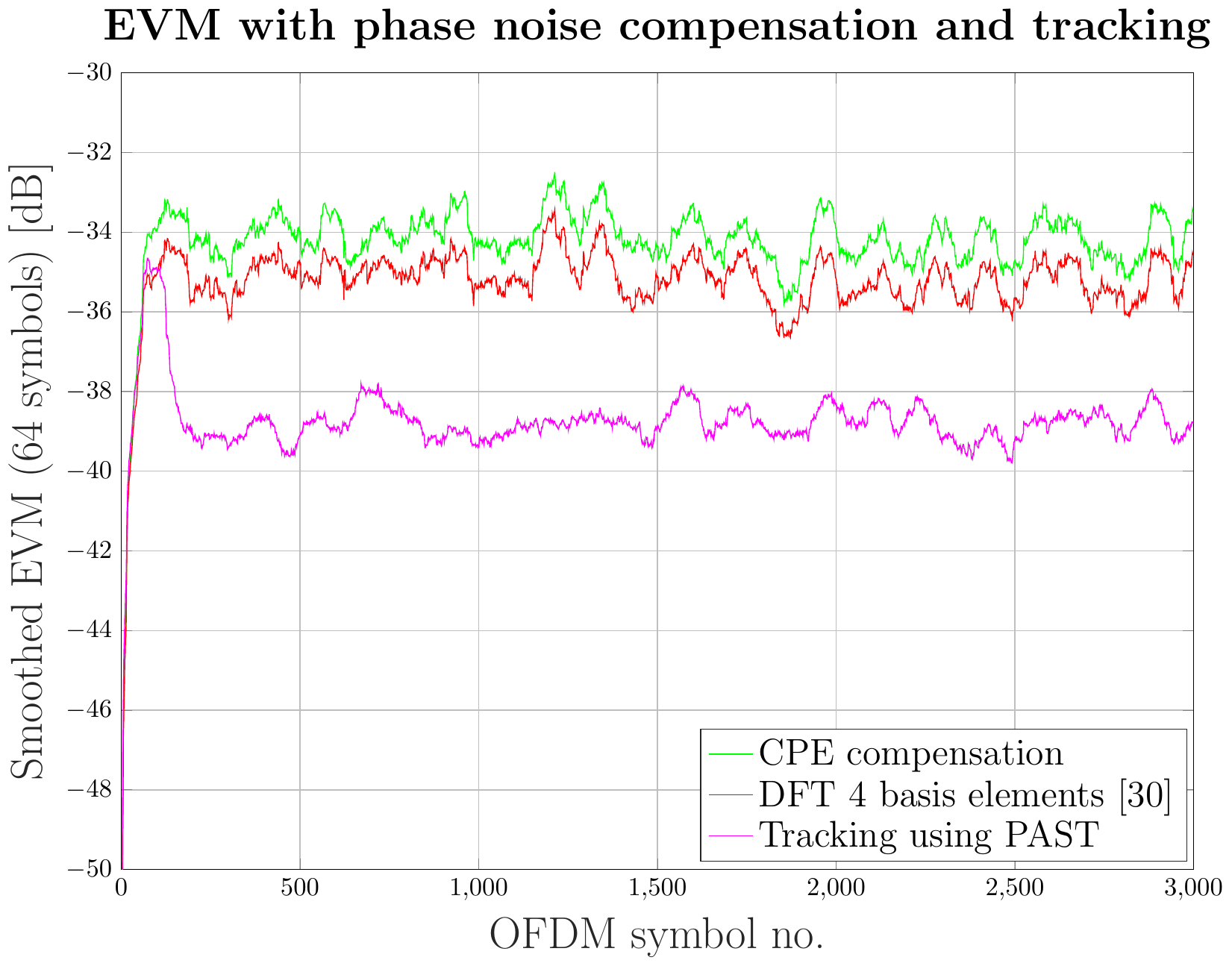}\\
  \caption{Limited training of 300 symbols. No residual carrier, $\sigma_{\phi}=3.5^{\circ }$ and $\beta =0.9$, one pair of receive channels}.
  \label{Figure15}
\end{figure}

\begin{figure}
  \centering
  % Requires \usepackage{graphicx}
  \includegraphics[scale=0.49]{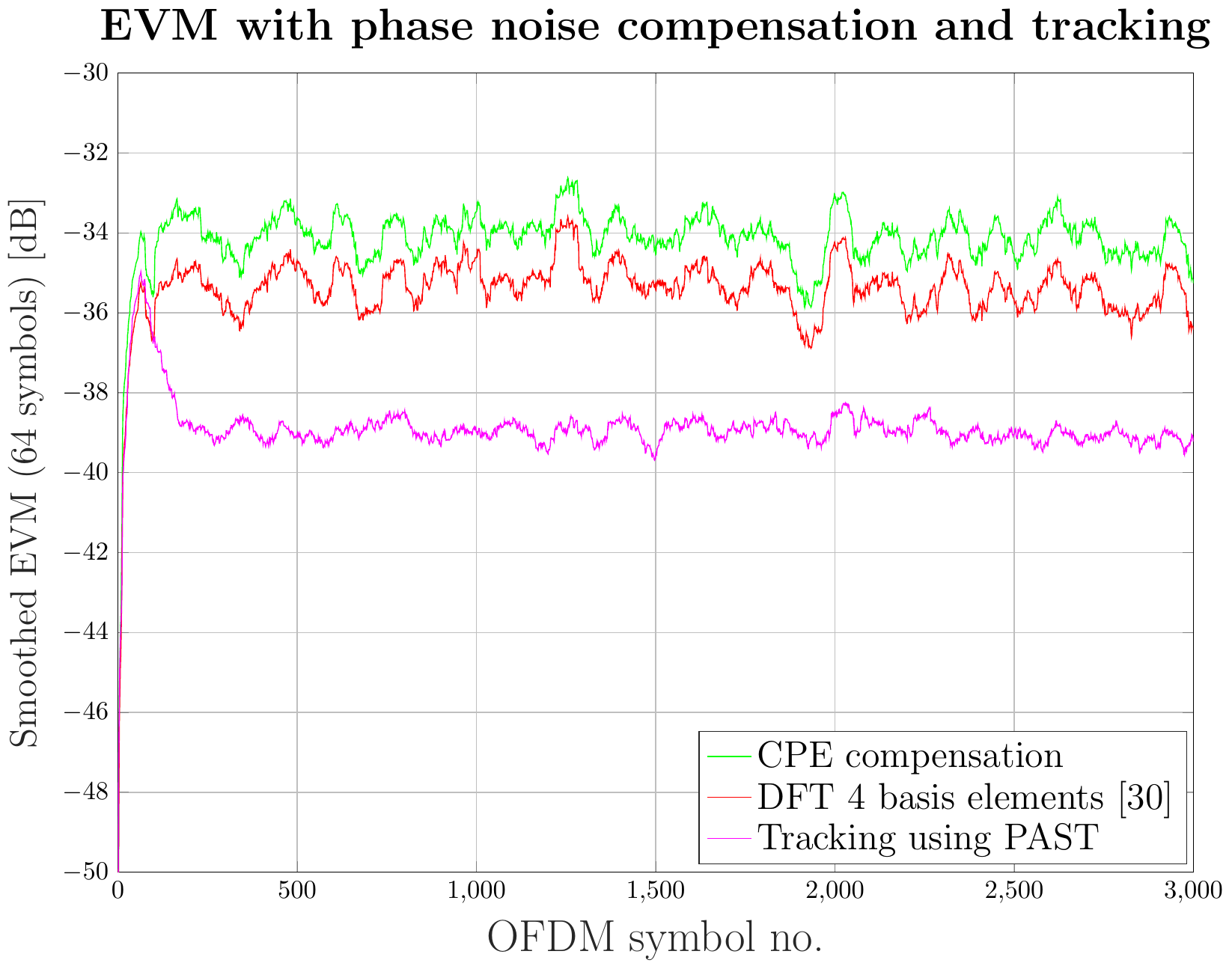}\\
  \caption{Limited training of 300 symbols. No residual carrier, $\sigma_{\phi}=3.5^{\circ }$ and $\beta =0.9$, averaged over 300 pairs of receive channels}.
  \label{Figure16}
\end{figure}

To test  the stationarity of the basis elements after convergence was achieved; we  repeated the experiment when no residual carrier was present and limited the training phase to $300$ symbols. This is important for the quality of the initial basis, based on previous estimates.
We performed two simulations whose results appear in Figure \ref{Figure15} and Figure \ref{Figure16}. Figure \ref{Figure15} depicts the results of the simulation for the two receive channels that are depicted in Figure \ref{Figure03}. Figure \ref{Figure16} depicts the results of the simulation averaged over 300 pairs of receive channels.
It is clear
 that  the basis vectors were relatively stationary since the EVM was fixed for the duration of the next $2700$ symbols.
 Further, the estimates were quite good and the mean EVM remained constant. This suggests that our stationarity assumption was sufficiently good. Note that both common phase removal and DFT based compensation were not as good and led to a $2-4\text{dB}$ loss.

\begin{figure}
  \centering
  % Requires \usepackage{graphicx}
  \includegraphics[scale=0.49]{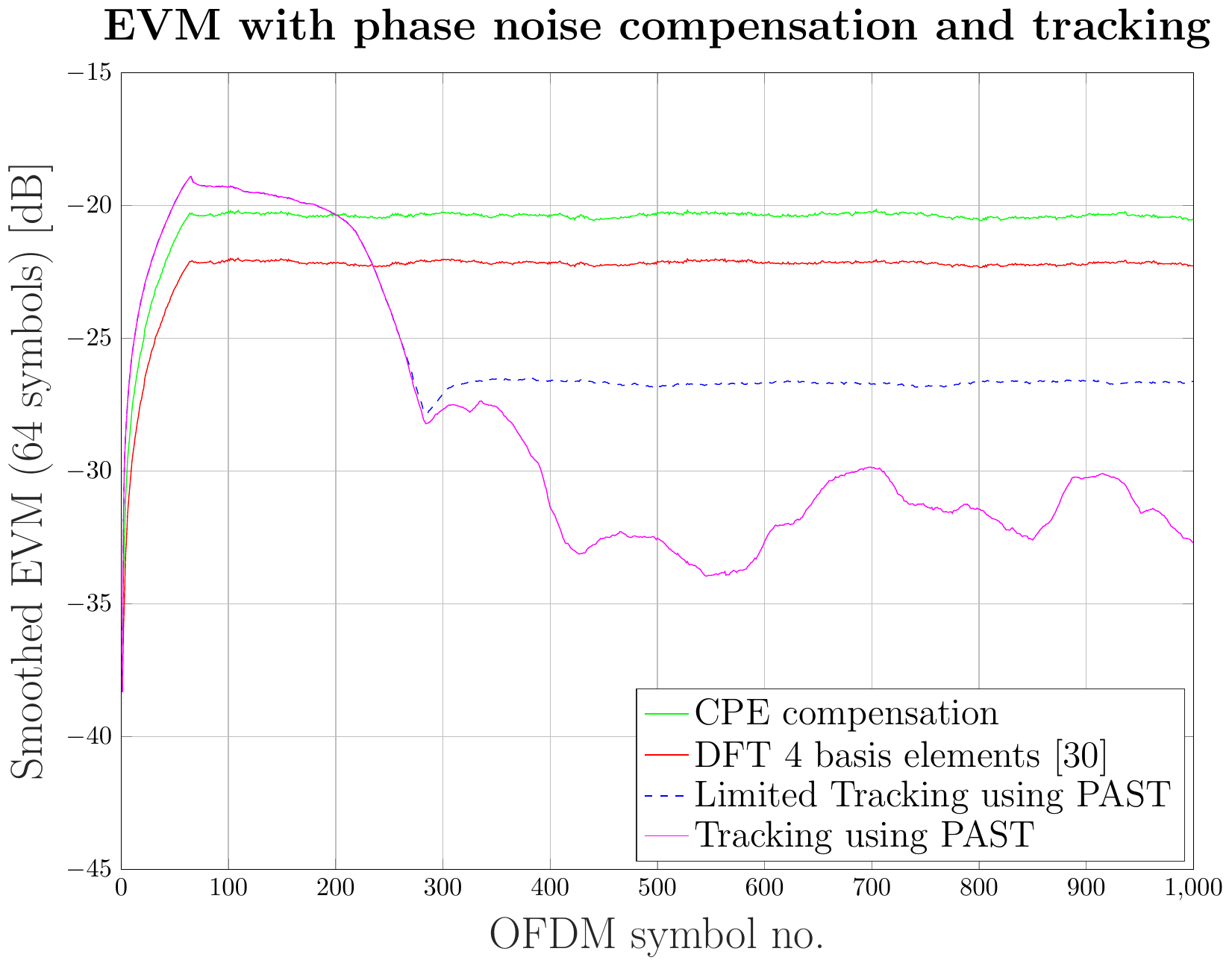}\\
  \caption{Comparison of the performance of phase noise compensation schemes with or without tracking.}
  \label{Figure17}
\end{figure}

To depict the advantages of tracking we performed the following simulation over the measured  channels, we set the residual carrier to be $5$ ppm and simulated phase noise of $3.5^{\circ}$, we averaged the results over 300 pairs of channels.
We stopped the tracking of the limited tracking  after 250 OFDM symbols and performed the tracking for the ``Tracking using PAST"  line during the whole simulation. As can be seen from Figure~\ref{Figure17}, tracking dramatically improves the behavior of the phase noise compensation scheme presented in this paper.

\section{Conclusion}\label{sec:conclisions}

We  presented a novel phase noise estimation technique. This technique leads to considerable reduction in phase noise and in particular works very well for strong phase noise. The specific basis proposed in this paper (using the eigenvectors of the noise correlation process) accounts primarily for its
 success compared to
 previous work. We  identified several possible extensions of the method to multi user MIMO systems as well as online calibration and exploitation of null tones. Finally, we  tested the possibility of decision directed tracking of the basis vectors. The tracking results using measured phase noise and residual carrier offset suggest that our stationarity assumption also holds.

\section*{Acknowledgement}
The authors would like to thank the anonymous reviewers and the associate editor
for their helpful and constructive comments which helped
improve the content of this paper.

\bibliographystyle{IEEETRAN}
%\bibliography{bib_phase}

   \begin{IEEEbiography}{Amir Leshem}
      (M'98-SM'06) received the B.Sc. degree (cum laude) in mathematics and physics and the M.Sc. (cum laude) and Ph.D. degrees in mathematics from the Hebrew University, Jerusalem,
Israel, in 1986, 1990, and 1998, respectively.
He is a Professor and one of the founders of the
Faculty of Engineering at Bar-Ilan University, Ramat-Gan, Israel, where he heads the Communications
track. He held visiting positions with Delft University, Delft, The Netherlands, and Stanford University,
Stanford, CA, USA, in 2008 to 2009. From 2000 to
2003, he was Director of Advanced Technologies with Metalink Broadband,
Yakum, Israel, where he was responsible for research and development of new
DSL and wireless MIMO modem technologies and served as a member of
ITU-T SG15, ETSI TM06, NIPP-NAI, and IEEE 802.3 and 802.11. From 2000
to 2002, he was also a Visiting Researcher with Delft University of Technology.
From 2003 to 2005, he was the Technical Manager of the U-BROAD consortium developing technologies to provide 100 Mb/s and beyond over copper
lines. From 1998 to 2000, he was with the Faculty of Information Technology
and Systems, Delft University of Technology, as a Postdoctoral Fellow working
on algorithms for the reduction of terrestrial electromagnetic interference in radio-astronomical radio-telescope antenna arrays and signal processing for communication. His main research interests include multichannel wireless and wireline communication, applications of game theory to dynamic and adaptive spectrum management of communication networks, software defined networks, array and statistical signal processing with applications to multiple-element
sensor arrays and networks, wireless communications, radio-astronomical imaging and brain research, Information processing in social networks, set theory, logic, and foundations of mathematics.

Prof. Leshem served as an Associate Editor for the IEEE TRANSACTIONS ON SIGNAL PROCESSING
from 2008 to 2011 and as a Guest Editor for several special issues of the \textit{IEEE Signal Processing Magazine} and the \textit{IIEEE JOURNAL OF SELECTED TOPICS IN SIGNAL PROCESSING}. Since 2010, he has been a member of the IEEE technical committee on Signal Processing for Communications and Networking.
    \end{IEEEbiography}

    \begin{IEEEbiography}{Michal Yemini}
received the B.Sc. in computer engineering from the Technion-Israel  Institute  of  Technology,  Haifa,  Israel, in 2011. She is currently completing her Ph.D. degree in the joint M.Sc.-Ph.D. program in the Faculty of Engineering, Bar-Ilan University, Ramat-Gan, Israel. Her main research interests include information theory, connectivity of large heterogeneous networks, dynamic spectrum allocation and resource management for wireless networks.
    \end{IEEEbiography}
\end{document}